\documentclass[preprint,nobibnotes,footinbib]{revtex4}%
\usepackage{amsfonts}
\usepackage{amsmath}
\usepackage{amssymb}
\usepackage{graphicx}%
\providecommand{\U}[1]{\protect\rule{.1in}{.1in}}

\begin{document}
\title{Physical Interpretation of Antigravity\thanks{Partly based on conference
lectures at \textquotedblleft String Field Theory-2015\textquotedblright%
\ (Sichuan Univ.), and \textquotedblleft Convergence\textquotedblright%
\ (Perimeter Inst.).}}
\thanks{Partly based on conference lectures at \textquotedblleft String Field
Theory-2015\textquotedblright, Sichuan University, China, and at
\textquotedblleft Convergence\textquotedblright, Perimeter Institute,
Waterloo, ON, Canada.}
\author{Itzhak Bars and Albin James}
\affiliation{Department of Physics and Astronomy, University of Southern California, Los
Angeles, CA 90089-0484, USA}
\pacs{PACS number}

\begin{abstract}
Geodesic incompleteness is a problem in both general relativity and string
theory. The Weyl invariant Standard Model coupled to General Relativity
(SM+GR), and a similar treatment of string theory, are improved theories that
are geodesically complete. A notable prediction of this approach is that there
must be antigravity regions of spacetime connected to gravity regions through
gravitational singularities such as those that occur in black holes and
cosmological bang/crunch. Antigravity regions introduce apparent problems of
ghosts that raise several questions of physical interpretation. It was shown
that unitarity is not violated but there may be an instability associated with
negative kinetic energies in the antigravity regions. In this paper we show
that the apparent problems can be resolved with the interpretation of the
theory from the perspective of observers strictly in the gravity region. Such
observers cannot experience the negative kinetic energy in antigravity
directly, but can only detect in and out signals that interact with the
antigravity region. This is no different than a spacetime black box\ for which
the information about its interior is encoded in scattering amplitudes for
in/out states at its exterior. Through examples we show that negative kinetic
energy in antigravity presents no problems of principles but is an interesting
topic for physical investigations of fundamental significance.

\end{abstract}
\maketitle
\tableofcontents

\section{Why Antigravity?}

The Lagrangian for the geodesically complete version of the Standard Model
coupled to General Relativity (SM+GR) is \cite{BST-conf}
\begin{equation}
\mathcal{L}\left(  x\right)  =\sqrt{-g}\left(
\begin{array}
[c]{c}%
L_{\text{SM}}\left(  A_{\mu}^{\gamma,W,Z,g},\;\psi_{q,l},\;\nu_{R}%
,\;\chi\right) \\
+g^{\mu\nu}\left(  \frac{1}{2}\partial_{\mu}\phi\partial_{\nu}\phi-D_{\mu
}H^{\dagger}D_{\nu}H\right) \\
-\left(  \frac{\lambda}{4}\left(  H^{\dagger}H-\omega^{2}\phi^{2}\right)
^{2}+\frac{\lambda^{\prime}}{4}\phi^{4}\right) \\
+\frac{1}{12}\left(  \phi^{2}-2H^{\dagger}H\right)  R\left(  g\right)
\end{array}
\right)  \label{action}%
\end{equation}
In the first line, $L_{SM}$ contains all the familiar degrees of freedom in
the properly extended conventional Standard Model, including gauge bosons
$(A_{\mu}^{\gamma,W,Z,g}),$ quarks \& leptons $\left(  \psi_{q,l}\right)  ,$
right-handed neutrinos $\nu_{R},$ dark matter $\chi,$ and their SU$\left(
3\right)  \times$SU$\left(  2\right)  \times$U$\left(  1\right)  $ gauge
invariant interactions among themselves and with the spin-0 fields $\left(
H,\phi\right)  $, where $H$=electroweak Higgs doublet, $\phi=$ a singlet. In
$L_{SM}$ all fields are minimally coupled to gravity. The second and third
lines describe the kinetic energy terms and interactions of the scalars among
themselves. The last term is the unique non-minimal coupling of conformal
scalars to the scalar curvature $R\left(  g\right)  $ that is required by
invariance of the full $\mathcal{L}\left(  x\right)  $ under local rescaling
(Weyl) with an arbitrary local parameter $\Omega\left(  x\right)  $
\begin{equation}%
\begin{array}
[c]{c}%
g_{\mu\nu}\rightarrow\Omega^{-2}g_{\mu\nu},~\phi\rightarrow\Omega
\phi,\;H\rightarrow\Omega H\\
\psi_{q,l}\rightarrow\Omega^{3/2}\psi_{q,l},\;A_{\mu}^{\gamma,W,Z,g}%
~\text{{\scriptsize unchanged.}}%
\end{array}
\end{equation}
If dark matter $\chi$ is a spin-0 field, then lines 2-4 in (\ref{action})
should be modified to treat $\chi$ as another conformally coupled scalar.

This theory has several pleasing features. There are no dimensionful
parameters, so all of those arise from a unique source, namely the gauge
fixing of the Weyl symmetry such as, $\phi\left(  x\right)  \rightarrow
\phi_{0},$ where $\phi_{0}$ is a dimensionful constant of the order of the
Planck scale. Then the gravitational constant is $\left(  16\pi G_{N}\right)
^{-1}=\phi_{0}^{2}/12$, the electroweak scale is $\langle\left\vert
H\right\vert \rangle=\omega\phi_{0},$ while dark energy, and masses for
quarks, leptons, gauge bosons, neutrinos and dark matter arise from
interactions with the scalars $\left(  \phi,H\right)  .$ The hierarchy of mass
scales is put in by hand through a hierarchy of dimensionless parameters. A
deeper theory is needed to explain this hierarchy, but in the present
effective theory it is at least possible to maintain it under renormalization
since dimensionless constants receive only logarithmic quantum corrections (no
need for low energy supersymmetry for the purpose of \textquotedblleft
naturalness\textquotedblright). To preserve the local scale symmetry in the
quantum theory one must adopt a Weyl invariant renormalization scheme in which
$\phi$ is the only renormalization scale, and consequently dimensionless
constants receive only Weyl invariant logarithmic renormalizations of the form
$\ln\left(  H/\phi\right)  ,$ etc. With such a renormalization scheme the
scale anomaly of all matter cancels against the scale anomaly of $\phi$
\cite{Percacci}, thus not spoiling the local symmetry. Then the unbroken Weyl
symmetry in the renormalized theory plays a central role in explaining the
smallness of dark energy as shown in \cite{IB-darkEnergy}. This also suggests
a definite relation between the electroweak vacuum and dark energy both of
which fill the entire universe.

The scalar $\phi$ is compensated by the Weyl symmetry, so $\phi$ is not a true
additional physical degree of freedom but, as a conformally coupled scalar,
participates in an important structure of the Weyl symmetry that has further
physical consequences involving antigravity spacetime regions in cosmology and
black holes as discussed in the following sections. The structure of interest,
that leads to the central discussion in the rest of this paper, is the
relative minus sign in $\left(  \phi^{2}-2H^{\dagger}H\right)  R$ and in the
scalar kinetic terms in (\ref{action}). These signs are compulsory and play an
important role in the geodesic completeness of the theory. With the given sign
patterns, $H$ has the correct sign for its kinetic term but $\phi$ has the
wrong sign. If $\phi$ had the same sign of kinetic energy as $H,$ then the
conformal coupling to $R$ would become purely negative which would lead to a
negative gravitational constant. So, to generate a positive gravitational
constant, $\phi$ must come with the opposite sign to $H.$ This makes $\phi$ a
ghost, but this is harmless since the Weyl symmetry can remove this ghost by a
gauge fixing.

This scheme has a straightforward generalization to supersymmetry/supergravity
and grand unification, but all scalars $\vec{s}$ must be conformally coupled,
$\left(  \phi^{2}-\vec{s}^{2}\right)  R,$ although some generalization is
permitted as long as the geodesically complete feature (related to signs) is
maintained \cite{BST-conf}. Furthermore, we point out that in all supergravity
theories, the curvature term has the form $\left(  1-K\left(  \varphi_{i}%
,\bar{\varphi}_{i}\right)  /3\right)  R,$ where $K$ is the K\"{a}hler
potential and $1$ represents the Einstein-Hilbert term \cite{Weinberg}. This
is again of the form $(\left\vert \phi\right\vert ^{2}-\left\vert \vec
{s}\right\vert ^{2})R$ with complex $\left(  \phi,\vec{s}\right)  ,$ where a
complex version of $\phi$ has been gauge fixed to 1 in a Weyl invariant
formulation of supergravity \cite{IB-sugra} (see also \cite{FerraraKallosh}).
Finally we emphasize that the same relative minus sign occurs also in a Weyl
invariant reformulation of low energy string theory (ST) but with a different
interpretation of $s$ related to the dilaton \cite{BST-string}. Hence the
structure $\left(  \phi^{2}-\vec{s}^{2}\right)  R$ is ubiquitous, but was
overlooked because it was commonly assumed that the gravitational constant, or
an effective structure that replaces it, could not or should not become
negative .

At the outset of this approach in 2008 \cite{IB-2Tgravity} the immediate
question was whether the dynamics would allow $\left(  \phi^{2}-s^{2}\right)
$ to remain always positive. It was eventually determined by Bars, Chen,
Steinhardt and Turok, in a series of papers during 2010-2012 (summary in
\cite{IB-cosmoSummary}), that the solutions of the field equations that do not
switch sign for this quantity are non-generic and of measure zero in the phase
space of initial conditions for the fields $\left(  \phi,s\right)  $. So,
according to the dynamics, it is untenable to insist on a limited patch
$\left\vert \phi\right\vert >\left\vert \vec{s}\right\vert $ of field space.
By contrast, it was found that the theory becomes geodesically complete when
all field configurations are included, thus solving generally the basic
problem of geodesic incompleteness.

The other side of the coin is that solving geodesic incompleteness comes with
the prediction that there would be antigravity sectors in the theory since the
effective gravitational constant that is proportional to $\left(  \phi
^{2}\left(  x\right)  -s^{2}\left(  x\right)  \right)  ^{-1}$ would
dynamically become negative in some spacetime regions. In view of the pleasing
features of the theory outlined in the second paragraph above, these
antigravity sectors must then be taken seriously and the corresponding new
physics must be understood. In our investigations so far, we discovered that
the antigravity sectors are geodesically connected to our own gravity sector
at gravitational singularities, like the big bang/crunch or black holes, which
occur precisely at the same spacetime points where $\left(  \phi^{2}\left(
x\right)  -s^{2}\left(  x\right)  \right)  $ vanishes or goes to infinity. The
related dynamical string tension \cite{BST-string}
\begin{equation}
T\left(  \phi,s\right)  \sim\left(  \phi+s\right)  ^{2\frac{1+\sqrt{d-1}}%
{d-2}}\left(  \phi-s\right)  ^{2\frac{1-\sqrt{d-1}}{d-2}},
\end{equation}
goes to zero or infinity simultaneously. So we need to figure out the physical
effects that can be observed in our universe due to the presence of
antigravity sectors behind cosmological \cite{BST-antigravity} and black hole
singularities \cite{ABJ-blackhole}. After overcoming several conceptual as
well as technical challenges we have been able to discuss some new physics
problems and developed new cosmological scenarios that involve an antigravity
period in the history of the universe \cite{BST-Higgs}\cite{turok}. A
remaining conceptual puzzle is an apparent possible instability in the
antigravity sector that is addressed and resolved in the remainder of this
paper. Our conclusion is that there are no fundamental problems but only
interesting physics of crucial significance.

\section{Geodesic completeness in the Einstein or string frames \label{frames}%
}

The classical or quantum analysis of this theory is best conducted in a Weyl
gauge we called the \textquotedblleft$\gamma$-gauge\textquotedblright%
\ \cite{BST-antigravity}\cite{BST-conf}\cite{BST-string} which amounts to
$\det\left(  -g\right)  =1.$ This allows the Sign$\left(  \phi^{2}%
-s^{2}\right)  $ to be determined by the dynamics. Note that the sign is gauge
invariant, so if the sign switches dynamically in one gauge it has to also
switch in all gauges. If one wishes to use the traditional \textquotedblleft
Einstein gauge\textquotedblright\ (E) or the \textquotedblleft string
gauge\textquotedblright\ (s) one can err by choosing an illegitimate gauge
that corresponds to a geodesically incomplete patch, such as\
\begin{equation}%
\begin{array}
[c]{l}%
\text{E}^{+}\text{-gauge:\ }\frac{1}{12}\left(  \phi_{E+}^{2}-s_{E+}%
^{2}\right)  =\frac{+1}{16\pi G_{N}},\\
\text{s}^{+}\text{-gauge:\ }\frac{d-2}{8\left(  d-1\right)  }\left(  \phi
_{s+}^{2}-s_{s+}^{2}\right)  =\frac{+1}{2\kappa_{d}^{2}}e^{-2\Phi}%
,\;\Phi=\text{dilaton.}%
\end{array}
\label{EsGuages}%
\end{equation}
The $E$ or $s$ subscripts on the fields indicate the gauge fixed form of the
corresponding field. If this were all, then there would be nothing new, and
the Weyl symmetry could be regarded as \textquotedblleft
fake\textquotedblright\ \cite{Jackiw}. However, the fact is that conventional
general relativity and string theory are geodesically incomplete because the
gauge choices just shown are valid only in the field patch in which
$\left\vert \phi\right\vert >\left\vert s\right\vert $. The dynamics
contradict the assumption of gauge fixing to only the positive patch. In the
negative regions one may choose again the Einstein or string gauge, but now
with a negative gravitational constant, $\frac{1}{12}\left(  \phi_{E-}%
^{2}-s_{E-}^{2}\right)  =\frac{-1}{16\pi G_{N}},$ or $\frac{d-2}{8\left(
d-1\right)  }\left(  \phi_{s-}^{2}-s_{s-}^{2}\right)  =\frac{-1}{2\kappa
_{d}^{2}}e^{-2\Phi}$. In those spacetime regions gravity is repulsive (antigravity).

The same situation arises in string theory. In the worldsheet formulation of
string theory the string tension is promoted to a background field $T\left(
\phi,s\right)  $ by connecting it directly to the features of the Weyl
invariant low energy string theory \cite{BST-string}. Then the string tension
$T\left(  \phi,s\right)  $ switches sign together with the corresponding
gravitational constant \cite{BST-string}. Thus the Weyl symmetric (SM+GR) and
ST predict that, in the Einstein or string gauges, one should expect a
\textit{sudden sign switch} of the effective Planck mass $\frac{1}{12}\left(
\phi^{2}-s^{2}\right)  $ at certain spacetime points that typically correspond
to singularities (e.g. big bang, black holes) encountered in the Einstein or
string frames.

One may choose better Weyl gauges (e.g. \textquotedblleft$\gamma
$-gauge\textquotedblright, choose $\det\left(  -g\right)  \rightarrow1$, or
\textquotedblleft$c$-gauge\textquotedblright, choose $\phi\rightarrow$
constant$)$ that cover globally all the positive and negative patches. Then
the sign switch of the effective Planck mass $\frac{1}{12}\left(  \phi
^{2}-s^{2}\right)  $ is smooth rather than abrupt.

However, if one wishes to work in the more familiar Einstein or string frames,
to recover the geodesically complete theory one must allow for the
gravitational constant to switch sign at singularities, and connect solutions
for fields across gravity/antigravity patches. In the $\pm$ Einstein gauges
shown above, the last term in Eq.(\ref{action}) becomes
\begin{equation}
\frac{\left(  \phi_{E\pm}^{2}-s_{E\pm}^{2}\right)  R\left(  g_{E\pm}\right)
}{12}=\frac{R\left(  g_{E\pm}\right)  }{\pm16\pi G_{N}}=\frac{R\left(  \pm
g_{E\pm}\right)  }{16\pi G_{N}}. \label{pmg}%
\end{equation}
where the $\pm$ for the gravity/antigravity regions can be absorbed into a
redefinition of the signature of the metric,
\begin{equation}
\hat{g}_{\mu\nu}^{E}=\pm g_{\mu\nu}^{E\pm}, \label{g-hat}%
\end{equation}
where the \textit{continuous} $\hat{g}_{\mu\nu}^{E}$ is the geometry in the
\textit{union} of the gravity/antigravity patches.

The same $\pm$ gauge choice is applied to every term in the SM+GR action in
Eq.(\ref{action}). Under the replacement $g_{\mu\nu}^{E-}\rightarrow-g_{\mu
\nu}^{E-}$ in the \textit{antigravity sector} some terms in the action flip
sign and some don't \cite{Duff}, e.g. $F_{\mu\nu}F^{\mu\nu}$ does not, but
$R\left(  g\right)  $ does as in Eq.(\ref{pmg}). One may be concerned that the
sign switches of the gravitational constant or the string tension may lead to
problems like unitarity or negative kinetic energy ghosts. We mention that
\cite{BST-string} has already argued that there are no unitarity problems due
to sign flips in field/string theories. There remains the question of possible
instability due to negative kinetic energy in the antigravity region. We show
in this paper that its presence is not a problem of principle for observers in
the gravity region and that those observers can detect interesting physical
effects related to the geodesically connected regions of antigravity.

\section{Unitarity and antigravity in cosmology}

There is a general impression that negative kinetic energy in field or string
theory implies ghosts associated with negative norm states. It is not
generally appreciated that negative norms (hence negative probabilities) are
automatically avoided by insisting on a strictly unitary quantization of the
theory. This has been illustrated in the quantization of the relativistic
harmonic oscillator \cite{IB-relativisticHO} with a timelike direction that
appears with the opposite sign to the spacelike directions, just like the
$\phi$ field as compared to the $H$ field in the SM+GR action in
Eq.(\ref{action}). Similar situations occur in the antigravity region where
some fields may appear with the wrong sign as described after Eq.(\ref{pmg}).
The first duty in quantization should be maintaining sanity in the meaning of
probability, as in \cite{IB-relativisticHO}, by avoiding a quantization
procedure that introduces negative norm states. Of course, there exist
successful cases, such as string theory in the \textquotedblleft covariant
quantization\textquotedblright\ procedure, that at first admits negative norms
to later kill them by applying constraints that select the positive norm
states. In principle, the relativistic oscillators in string theory could also
be treated as in \cite{IB-relativisticHO} and very likely still recover the
same gauge invariant physical states without ever introducing negative norm
states in string theory. It would be preferable to quantize without negative
norm states at all from the very beginning.

When there is not enough gauge symmetry to remove a degree of freedom that has
the wrong sign of kinetic energy, a unitary quantization procedure like
\cite{IB-relativisticHO} maintains unitarity. However, the effect of the
negative kinetic energy is to cause an instability (not unlike a tachyonic
mass term, or a bottomless potential, would), so that there may not be a
ground state for that degree of freedom while it propagates in the antigravity
region. This is the negative kinetic energy issue in the antigravity sector.
Perhaps some complete theory as a whole conspires to have a ground state even
in antigravity. Although this would be reassuring, it appears that this is not
necessary in order to make sense of the physics as detected by observers in
the gravity sector. Such observers can verify that the same degree of freedom
does have a ground state in the gravity region while they can never experience
directly the negative kinetic energy in the antigravity sector. The only
physics questions that make sense for those observers is what can be learned
about the existence of antigravity through scattering experiments that involve
in/out states as defined in the gravity region. For those questions the issue
of whether there is a ground state in the antigravity region does not matter,
but unitarity continues to matter. Therefore we point out how this works in
the case of cosmology that admits an antigravity region.

\subsection{WdW equation and unitarity in mini superspace}

The Wheeler de Witt (WdW) equation is the quantum version of the $\mu=0$ and
$\nu=0$ component of the Einstein equation, $\left(  G_{00}-T_{00}\right)
\psi=0.$ This is a constraint applied on physical states in covariant
quantization of general relativity \cite{deWitt}. The \textquotedblleft mini
superspace\textquotedblright\ consists of only time dependent (homogeneous)
scalar fields $\left(  \phi\left(  x^{0}\right)  ,s\left(  x^{0}\right)
\right)  $ and the FRW metric, $ds^{2}=a^{2}\left(  x^{0}\right)  \left(
-\left(  dx^{0}\right)  ^{2}+\gamma_{ij}\left(  x^{0},\vec{x}\right)
dx^{i}dx^{j}\right)  ,$ with $\gamma_{ij}$ describing spacial curvature and
anisotropies, while $T_{00}$ includes the radiation density, $\rho_{r}\left(
x^{0}\right)  /a^{4}\left(  x^{0}\right)  .$ From the action in
Eq.(\ref{action}) we can derive a Wheeler de Witt equation that is invariant
under Weyl rescalings $\left(  \phi,s,a\right)  \rightarrow\left(  \Omega
\phi,\Omega s,\Omega^{-1}a\right)  $ with a time dependent $\Omega\left(
x^{0}\right)  $; this allows us to choose a gauge. To allow $\left(  \phi
^{2}-s^{2}\right)  $ to have any sign dynamically, we prefer the $\gamma
$-gauge given by
\begin{equation}
\left(  \phi,s,a\right)  \rightarrow\left(  \phi_{\gamma},s_{\gamma},1\right)
,\text{ or }a_{\gamma}\left(  x^{0}\right)  =1.
\end{equation}
We concentrate here on the simplest FRW geometry in the $\gamma$-gauge,
\begin{equation}
ds_{\gamma}^{2}=-\left(  dx^{0}\right)  ^{2}+\frac{dr^{2}}{1-Kr^{2}}%
+r^{2}d\Omega^{2}, \label{K}%
\end{equation}
with no anisotropy or inhomogeneities, but with a positive constant spatial
curvature $K>0$. This is not realistic, but it is the easiest case to
illustrate the unitarity properties of the quantum theory that includes
antigravity regions (more degrees of freedom, and negative or zero $K$ would
be treated in a similar manner). The mini superspace is just $\left(
\phi_{\gamma},s_{\gamma}\right)  ,$ while the constraint $\left(
T_{00}-G_{00}\right)  =0$ derived from (\ref{action}) is, $-\frac{1}{2}%
\dot{\phi}_{\gamma}^{2}+\frac{1}{2}\dot{s}_{\gamma}^{2}+\frac{1}{2}K\left(
-\phi_{\gamma}^{2}+s_{\gamma}^{2}\right)  +\rho_{r}=0.$ This is recognized as
the Hamiltonian for the relativistic harmonic oscillator, $H=\frac{1}%
{2}\left(  \dot{x}^{2}+Kx^{2}\right)  ,$ with $x^{\mu}\left(  \tau\right)
=\left(  \phi_{\gamma}\left(  \tau\right)  ,s_{\gamma}\left(  \tau\right)
\right)  ,$ subject to the constraint, $H+\rho_{r}=0,$ where $\rho_{r}$ is a
constant. Note that this Hamiltonian contains negative energy for the
(time-like) $\phi_{\gamma}$ degree of freedom. Recall that we have already
used up the Weyl symmetry so this degree of freedom cannot be removed and its
negative energy must be dealt with. The naive quantization of the relativistic
harmonic oscillator would introduce negative norm states for the $\phi
_{\gamma}$ degree of freedom (as in string theory), so it appears there may be
trouble with unitarity. However, this is not the case, because this system
(and similar cases) can be quantized by respecting unitarity without ever
introducing negative norms as shown in \cite{IB-relativisticHO}. This goes as
follows: the quantum system obeys the constraint equation $\left(  H+\rho
_{r}\right)  \Psi=0.$ This is the WdW equation that takes the form%
\begin{equation}
\left(  \frac{1}{2}\partial_{\phi_{\gamma}}^{2}-\frac{1}{2}\partial
_{s_{\gamma}}^{2}+\frac{K}{2}\left(  -\phi_{\gamma}^{2}+s_{\gamma}^{2}\right)
+\rho_{r}\right)  \Psi\left(  \phi_{\gamma},s_{\gamma}\right)  =0. \label{WdW}%
\end{equation}
This is recognized as the Klein-Gordon equation for the quantized relativistic
harmonic oscillator. The eigenstates and eigenvalues of the independent
$\phi_{\gamma}$ and $s_{\gamma}$ oscillators are
\begin{equation}%
\begin{array}
[c]{l}%
\frac{1}{2}\left(  -\partial_{\phi_{\gamma}}^{2}+K\phi_{\gamma}^{2}\right)
\psi_{n_{\phi}}\left(  \phi_{\gamma}\right)  =\sqrt{K}\left(  n_{\phi}%
+\frac{1}{2}\right)  \psi_{n_{\phi}}\left(  \phi_{\gamma}\right)  ,\\
\frac{1}{2}\left(  -\partial_{s_{\gamma}}^{2}+Ks_{\gamma}^{2}\right)
\psi_{n_{s}}\left(  s_{\gamma}\right)  =\sqrt{K}\left(  n_{s}+\frac{1}%
{2}\right)  \psi_{n_{s}}\left(  s_{\gamma}\right)  ,
\end{array}
\end{equation}
where $\left(  n_{\phi},n_{s}\right)  $ are positive integers, $0,1,2,3,\cdots
,$ and the explicit \textit{positive norm} complete set of off-shell solutions
are%
\begin{equation}%
\begin{array}
[c]{c}%
\Psi_{n_{\phi}n_{s}}\left(  \phi_{\gamma},s_{\gamma}\right)  =\psi_{n_{\phi}%
}\left(  \phi_{\gamma}\right)  \psi_{n_{s}}\left(  s_{\gamma}\right)  ,\\
\psi_{n_{\phi}}\left(  \phi_{\gamma}\right)  =A_{n_{\phi}}e^{-\frac{1}{2}%
\sqrt{K}\phi_{\gamma}^{2}}H_{n_{\phi}}\left(  \phi_{\gamma}\right)  ,\\
\psi_{n_{s}}\left(  s_{\gamma}\right)  =A_{n_{s}}e^{-\frac{1}{2}\sqrt
{K}s_{\gamma}^{2}}H_{n_{s}}\left(  s_{\gamma}\right)  ,
\end{array}
\label{complete}%
\end{equation}
where $H_{n}\left(  z\right)  $ are the Hermite polynomials and $A_{n_{\phi}%
},A_{n_{s}}$ are normalization constants. Then the WdW equation (\ref{WdW}) is
solved by constraining the eigenvalues, $\sqrt{K}\left(  -n_{\phi}%
+n_{s}\right)  +\rho_{r}=0.$ Hence the complete on-shell basis that satisfies
the constraint is
\begin{equation}
\Psi_{n}\left(  \phi_{\gamma},s_{\gamma}\right)  =A_{n+r}A_{n}e^{-\frac
{\sqrt{K}}{2}\left(  \phi_{\gamma}^{2}+s_{\gamma}^{2}\right)  }H_{n+r}\left(
\phi_{\gamma}\right)  H_{n}\left(  s_{\gamma}\right)  , \label{RelHO}%
\end{equation}
with $n=0,1,2,\cdots,$ where we defined
\begin{equation}
n_{s}\equiv n,\text{ \ }n_{\phi}\equiv n+r,\;\text{ and }\frac{\rho_{r}}%
{\sqrt{K}}\equiv r\;\text{a fixed integer.}%
\end{equation}
If $\frac{\rho_{r}}{\sqrt{K}}$ is not an integer there is no solution to the
constraint, hence radiation must be quantized for this system to be
non-trivial at the quantum level. The general on-shell solution of the WdW
equation is an arbitrary superposition of this basis%
\begin{equation}
\Psi\left(  \phi_{\gamma},s_{\gamma}\right)  =\sum_{n=0}^{\infty}c_{n}\Psi
_{n}\left(  \phi_{\gamma},s_{\gamma}\right)
\end{equation}
The complex coefficients $c_{n}$ are chosen to insure that $\Psi\left(
\phi_{\gamma},s_{\gamma}\right)  $ is normalized.

All quantum states have positive norm and unitarity is satisfied. $\Psi\left(
\phi_{\gamma},s_{\gamma}\right)  $ is the probability amplitude for where the
system is in the $\left(  \phi_{\gamma},s_{\gamma}\right)  $ plane. The
gravity/antigravity regions are $\phi_{\gamma}^{2}\lessgtr s_{\gamma}^{2}.$
Evidently there is no way of preventing the generic wavefunctions from being
non-zero in the antigravity region, so the system generically evolves through
both the gravity and antigravity regions.

We emphasize that the quantization method in \cite{IB-relativisticHO} that we
used to maintain unitarity is very different than the quantization of the
relativistic oscillator used in string theory. In string theory one defines
relativistic creation/annihilation operators $a_{\mu},a_{\mu}^{\dagger}$ and a
vacuum state that satisfies $a_{\mu}|0\rangle=0.$ Then the quantum states at
level $l$ are given by applying $l$ creation operators, $a_{\mu_{1}}^{\dagger
}a_{\mu_{2}}^{\dagger}\cdots a_{\mu_{l}}^{\dagger}|0\rangle.$ The vacuum state
is Lorentz invariant, while the states at level $l$ form a collection of
\textit{finite dimensional} irreducible representations of the Lorentz group.
All the states at level $l$ have \textit{positive energy}, $E_{l}=\sqrt
{K}\left(  l+1\right)  .$ The constraint $H+\rho_{r}=0$ (WdW equation) can be
satisfied only for negative quantized $\rho_{r}$ at only one level
$l=-1+\left\vert \rho_{r}\right\vert /\sqrt{K}.$ In position space the vacuum
state takes the Lorentz invariant form $\psi_{0}\left(  x^{\mu}\right)  \sim
e^{-\sqrt{K}x^{2}}=e^{-\sqrt{K}\left(  -\phi_{\gamma}^{2}+s^{2}\right)  }$,
while the the states at level $l$ are of the form of a polynomial of $x^{\mu}$
of degree $l$ multiplied by the same exponential $e^{-\sqrt{K}x^{2}}.$ A
subset of the level-$l$ states have negative norm because finite dimensional
representations are not unitary representations of the Lorentz group, so this
method of quantization gets into trouble with unitarity. We contrast this
result to ours in Eq.(\ref{RelHO}) where we have displayed an infinite, rather
than finite, number of states and a Gaussian factor $e^{-\sqrt{K}\left(
\phi_{\gamma}^{2}+s^{2}\right)  }$ that converges in all directions, rather
than the non-convergent Lorentz invariant form $e^{-\sqrt{K}\left(
-\phi_{\gamma}^{2}+s^{2}\right)  }.$ There is no Lorentz invariant vacuum
state. As shown in \cite{IB-relativisticHO}, our states in Eq.(\ref{RelHO})
form an \textit{infinite dimensional unitary representation} of the Lorentz
group for which all the states have positive norm. Furthermore, those that
satisfy the constraint have positive total energy, $H=\rho_{r},$ as long as
$\rho_{r}$ is positive. However, as seen in Eq.(\ref{complete}), there are
\textit{off-shell states} of positive as well as negative energy. These
remarks make it clear that the price for maintaining unitarity (which is the
first duty in quantization) is the presence of regions of spacetime with
negative kinetic energy, which, in our case, amounts to regions of
antigravity. Our task in this paper is to explain that negative kinetic energy
in the antigravity sector does not necessarily imply a problem by interpreting
the physical significance of antigravity.

\subsection{Feynman propagator in mini superspace}

The Feynman propagator associated with this WdW equation is%
\begin{equation}
G\left(  \phi^{\prime},s^{\prime};\phi,s\right)  =\langle\phi^{\prime
},s^{\prime}|\frac{i}{H+\rho_{r}+i\varepsilon}|\phi,s\rangle.
\end{equation}
We can use the complete basis $|n_{\phi},n_{s}\rangle$ to insert identity in
terms of the eigenstates of the off-shell $H=-H_{\phi}+H_{s}$ operator without
any constraints on the integers $\left(  n_{\phi},n_{s}\right)  .$ Then we
compute%
\begin{equation}%
\begin{array}
[c]{l}%
G\left(  \phi^{\prime},s^{\prime};\phi,s\right)  =i\sum_{n_{\phi},n_{s}\geq
0}\frac{\langle\phi^{\prime},s^{\prime}|n_{\phi},n_{s}\rangle\langle n_{\phi
},n_{s}|\phi,s\rangle}{-n_{\phi}+n_{s}+\rho_{r}+i\varepsilon}\\
\;\;=i\sum_{n_{\phi},n_{s}\geq0}\psi_{n_{\phi}}\left(  \phi^{\prime}\right)
\psi_{n_{s}}\left(  s^{\prime}\right)  \psi_{n_{\phi}}^{\ast}\left(
\phi\right)  \psi_{n_{s}}^{\ast}\left(  s\right)  \left(  -n_{\phi}+n_{s}%
+\rho_{r}+i\varepsilon\right)  ^{-1}\\
\;\;=i\sum_{n_{\phi},n_{s}\geq0}\int_{0}^{\infty}d\tau~\psi_{n_{\phi}}\left(
\phi^{\prime}\right)  \psi_{n_{s}}\left(  s^{\prime}\right)  \psi_{n_{\phi}%
}^{\ast}\left(  \phi\right)  \psi_{n_{s}}^{\ast}\left(  s\right)  \left(
-i~e^{i\tau\left(  -n_{\phi}+n_{s}+\rho_{r}+i\varepsilon\right)  }\right) \\
\;\;=\int_{0}^{\infty}d\tau e^{i\tau\left(  \rho_{r}+i\varepsilon\right)
}\langle\phi^{\prime}|e^{-i\tau H_{\phi}}|\phi\rangle~\langle s^{\prime
}|e^{i\tau H_{s}}|s\rangle\\
\;\;=\int_{0}^{\infty}d\tau\frac{\sqrt{K}~e^{i\tau\left(  \rho_{r}%
+i\varepsilon\right)  }}{2\pi\sin\left(  \sqrt{K}\tau\right)  }\exp\left(
\frac{-i\sqrt{K}}{2\sin\left(  \sqrt{K}\tau\right)  }\left[  \left(
x^{2}+x^{\prime2}\right)  \cos\sqrt{K}\tau-2x\cdot x^{\prime}\right]  \right)
\end{array}
\label{Feynman}%
\end{equation}
In the last step we used the propagator $\langle\phi^{\prime}|e^{-i\tau
H_{\phi}}|\phi\rangle$ for the $1$-dimensional harmonic oscillator, and then
substituted $x^{2}=-\phi^{2}+s^{2}$ and $x\cdot x^{\prime}=-\phi\phi^{\prime
}+ss^{\prime}.$ This quantum computation in the Hamiltonian formalism agrees
with the path integral computation in \cite{turok}.

The Feynman propagator is a measure of the probability that the system that
starts in some initial state will be found in some final state. For observers
outside of the antigravity region the initial and final states $|\phi
,s\rangle,|\phi^{\prime},s^{\prime}\rangle$ are both in the gravity region,
$\left\vert \phi\right\vert >\left\vert s\right\vert $ and $\left\vert
\phi^{\prime}\right\vert >\left\vert s^{\prime}\right\vert ,$ although during
the propagation from initial to final state the antigravity region is probed
as seen from the sums over $\left(  n_{\phi},n_{s}\right)  $ where both
positive and negative energy states of the off-shell Hamiltonian $H=-H_{\phi
}+H_{s}$ enter in the calculation. We see from the last expression in
Eq.(\ref{Feynman}) that $G\left(  \phi^{\prime},s^{\prime};\phi,s\right)  $ is
a perfectly reasonable function indicating that there are no issues with
fundamental principles in this calculation which involves an intermediate
period of antigravity in the evolution of the universe.

This was the case of a radiation dominated spatially curved spacetime which is
far from being a generic configuration in the early universe close to the
singularity. The generic dominant terms in the Einstein frame are the kinetic
energy of the scalar and anisotropy (in the spatial metric) and the next
non-leading term is radiation. The sub-dominant terms, including curvature,
inhomogeneities, potential energy, dark energy, etc. are negligible near the
singularity. The dominant generic behavior near the singularity was computed
classically in \cite{BST-antigravity} where it was discovered that there
\textit{must} be an inescapable excursion into the antigravity regime before
coming back to the gravity sector, as outlined in the previous paragraph.
Hence, a similar computation to Eq.(\ref{Feynman}), by using the dominant
terms in the WdW equation (instead of (\ref{WdW})) should replace our
computation here. Unpublished work along these lines dating back to 2011
\cite{IB-cosmoSummary} indicate that the physical picture already obtained
through classical solutions in \cite{BST-antigravity} continues to hold in
mini-superspace at the quantum level.

\subsection{More general WdW equation}

As mentioned in the previous paragraph, we can generalize the WdW equation
above (\ref{WdW}) by including the physical features that would make it more
realistic for a description of the early universe in terms of a mini
superspace. This includes the kinetic energy for anisotropy degrees of freedom
that cannot be neglected when $s^{2}/\phi^{2}\simeq1$ close to the
singularity,  and the potential energy terms for both the scalars and
anisotropies that tend to become important when $\left\vert 1-s^{2}/\phi
^{2}\right\vert \gtrsim1.$ The action for the mini-superspace that includes
these features is given in Eq.(8) in \cite{IB-cosmoSummary}. The corresponding
WdW equation in the $\gamma$-gauge modifies Eq.(\ref{WdW}) as follows
\begin{equation}
\left(
\begin{array}
[c]{c}%
\frac{1}{2}\partial_{\phi_{\gamma}}^{2}-\frac{1}{2}\partial_{s_{\gamma}}%
^{2}-\frac{1}{2}\frac{1}{\phi_{\gamma}^{2}-s_{\gamma}^{2}}\left(
\partial_{\alpha_{1}}^{2}+\partial_{\alpha_{2}}^{2}\right)  +\rho_{r}\\
+V\left(  \phi_{\gamma},s_{\gamma}\right)  -\frac{1}{2}\left(  \phi_{\gamma
}^{2}-s_{\gamma}^{2}\right)  v\left(  \alpha_{1},\alpha_{2}\right)
\end{array}
\right)  \Psi\left(  \phi_{\gamma},s_{\gamma},\alpha_{1},\alpha_{2}\right)
=0.
\end{equation}
The additional anisotropy degrees of freedom $\left(  \alpha_{1},\alpha
_{2}\right)  $ are part of the 3-dimensional metrics of types Kasner,
Bondi-VIII, and Bondi-IX, as shown in Eq.(7) of Ref.\cite{IB-cosmoSummary}.
\ The term containing the anisotropy potential $v\left(  \alpha_{1},\alpha
_{2}\right)  $ above vanishes for the Kasner metric, while for Bondi-VIII and
IX \ it simplifies to a constant, $v\left(  \alpha_{1},\alpha_{2}\right)
\rightarrow K,$ in the zero anisotropy limit (as in Eq.(\ref{WdW})). The
details of the anisotropy potential $v\left(  \alpha_{1},\alpha_{2}\right)  $
are given in Eq.(9) of Ref.\cite{IB-cosmoSummary}.

With these additional features the WdW equation is no longer separable in the
$\left(  \phi_{\gamma},s_{\gamma}\right)  $ degrees of freedom. To make
progress we make a change of variables by defining
\begin{equation}
z=\phi_{\gamma}^{2}-s_{\gamma}^{2},\;\sigma=\frac{1}{2}\ln\left\vert
\frac{\phi_{\gamma}+s_{\gamma}}{\phi_{\gamma}-s_{\gamma}}\right\vert .
\end{equation}
Note that $z\gtrless0$ corresponds to gravity/antigravity. These variables
were used in the classical analysis of the same system in
\cite{BST-antigravity} where the classical equations that follow from the same
action were studied. Weyl invariance requires $V\left(  \phi,s\right)  $ to be
a homogeneous function of degree four, $V\left(  t\phi,ts\right)
=t^{4}V\left(  \phi,s\right)  ,$ so without loss of generality we may write
\begin{equation}
V\left(  \phi_{\gamma},s_{\gamma}\right)  =z^{2}v\left(  \sigma\right)  ,
\end{equation}
where $v\left(  \sigma\right)  $ is any function that is specified by some
physical model. The WdW equation above takes the following form in the
$z,\sigma,\alpha_{1},\alpha_{2}$ basis
\begin{equation}
\left(
\begin{array}
[c]{c}%
\partial_{z}^{2}+\frac{1}{4z^{2}}\left(  -\partial_{1}^{2}-\partial_{2}%
^{2}-\partial_{\sigma}^{2}+1\right)  \\
+\frac{z}{2}v\left(  \sigma\right)  -\frac{1}{4}v\left(  \alpha_{1},\alpha
_{2}\right)  +\frac{\rho_{r}}{2z}%
\end{array}
\right)  \left(  z^{1/2}\Psi\left(  z,\sigma,\alpha_{1},\alpha_{2}\right)
\right)  =0.\label{TotPsi}%
\end{equation}
Near the singularity, $z=0,$ assuming the potentials are neglected compared to
the dominant and subdominant $z^{-2},z^{-1}$ terms, the wavefunction becomes
separable in the form of a 3-dimensional plane wave, $\Psi=\exp\left(
ip_{1}\alpha_{1}+ip_{2}\alpha_{2}+ip_{3}\sigma\right)  \psi\left(  z\right)  $
with constant \textquotedblleft momenta\textquotedblright\ $\left(
p_{1},p_{2},p_{3}\right)  ,$ thus reducing (\ref{TotPsi}) to an ordinary
second order differential equation
\begin{equation}
\left(  \partial_{z}^{2}+\frac{1}{4z^{2}}\left(  p_{1}^{2}+p_{2}^{2}+p_{3}%
^{2}+1\right)  +\frac{\rho_{r}}{2z}\right)  \left(  z^{1/2}\psi\left(
z\right)  \right)  \simeq0.\label{closeToZero}%
\end{equation}
This is recognized as a Hydrogen-atom type differential equation; its
solutions are given analytically in terms of special functions related to the
representations of SL$\left(  2,R\right)  $ as discussed in \cite{IB-QM}. From
these wavefunctions we learn that the behavior of the probability
distributions near the singularity, $z\sim0$ where the gravity/antigravity
transition occurs, matches closely the behavior of the unique analytic
classical \textquotedblleft attractor\textquotedblright\ solution that
corresponds to the \textquotedblleft antigravity loop\textquotedblright%
\ described in \cite{BST-antigravity}. Namely, with a non-zero parameter,
$p\equiv\sqrt{p_{1}^{2}+p_{2}^{2}+p_{3}^{2}},$ the cosmological evolution of
the universe cannot avoid to pass temporarily though an antigravity sector,
$z<0$. Meanwhile, as seen here, the wavefunctions are normalizable and fully
consistent with unitarity in both the gravity and antigravity sectors
$z\gtrless0$. This conclusion, in the presence of the dominant anisotropy
terms, is in agreement with the lessons learned above with the simpler form of
the WdW equation in (\ref{WdW}).

As $\left\vert z\right\vert $ increases beyond the singularity and reaches
$\left\vert z\right\vert \sim1$, in either the gravity or antigravity
sectors,  the potentials $v\left(  \sigma\right)  $ and $v\left(  \alpha
_{1},\alpha_{2}\right)  $ can no longer be neglected. We may still reduce the
4-variable partial differential equation to a single-variable ordinary
differential equation as follows. Define the wavefunctions $\Phi_{n}\left(
\sigma\right)  $ and $\xi_{m_{1}m_{2}}\left(  \alpha_{1},\alpha_{2}\right)  $
as follows
\begin{equation}%
\begin{array}
[c]{c}%
\left(  -\partial_{\sigma}^{2}+2z^{3}v\left(  \sigma\right)  \right)  \Phi
_{n}\left(  \sigma\right)  =E_{n}\left(  z\right)  \Phi_{n}\left(
\sigma\right)  \\
\left(  -\partial_{1}^{2}-\partial_{2}^{2}-z^{2}v\left(  \alpha_{1},\alpha
_{2}\right)  \right)  \xi_{m_{1}m_{2}}\left(  \alpha_{1},\alpha_{2}\right)
=E_{m_{1}m_{2}}\left(  z\right)  \xi_{m_{1}m_{2}}\left(  \alpha_{1},\alpha
_{2}\right)
\end{array}
\label{PhiXi}%
\end{equation}
In solving these equations the parameter $z$ is considered a constant
parameter, but the eigenvalues $E_{n}\left(  z\right)  $ and $E_{m_{1}m_{2}%
}\left(  z\right)  $ clearly depend on $z.$ Then, writing the wavefunction in
separable form,
\begin{equation}
\Psi\left(  z,\sigma,\alpha_{1},\alpha_{2}\right)  \sim\Psi_{n,m_{1},m_{2}%
}\left(  z\right)  \times\Phi_{n}\left(  \sigma\right)  \times\xi_{m_{1}m_{2}%
}\left(  \alpha_{1},\alpha_{2}\right)  ,
\end{equation}
Eq.(\ref{TotPsi}) reduces to
\begin{equation}
\left(  \partial_{z}^{2}+\frac{1}{4z^{2}}\left(  E_{n}\left(  z\right)
+E_{m_{1}m_{2}}\left(  z\right)  +1\right)  +\frac{\rho_{r}}{2z}\right)
\left(  z^{1/2}\Psi_{n,m_{1},m_{2}}\left(  z\right)  \right)  =0.\label{zEq}%
\end{equation}
while the general solution takes the form
\begin{equation}
\Psi\left(  z,\sigma,\alpha_{1},\alpha_{2}\right)  =\sum_{n,m_{1},m_{2}%
}c_{n,m_{1},m_{2}}\Psi_{n,m_{1},m_{2}}\left(  z\right)  \times\Phi_{n}\left(
\sigma\right)  \times\xi_{m_{1}m_{2}}\left(  \alpha_{1},\alpha_{2}\right)
\end{equation}
with arbitrary constant coefficients $c_{n,m_{1},m_{2}}.$

As we saw above, when $v\left(  \sigma\right)  ,v\left(  \alpha_{1},\alpha
_{2}\right)  $ are zero, the corresponding energies tend to constants
$E_{n}\left(  z\right)  \rightarrow p_{3}^{2}$ and $E_{m_{1}m_{2}}\left(
z\right)  \rightarrow p_{1}^{2}+p_{2}^{2},$ so this can be used as a guide to
the role played by $E_{n}\left(  z\right)  $ and $E_{m_{1}m_{2}}\left(
z\right)  $ in Eq.(\ref{zEq}). Using simple models for $v\left(
\sigma\right)  ,v\left(  \alpha_{1},\alpha_{2}\right)  $ to extract properties
of $E_{n}\left(  z\right)  $ and $E_{m_{1}m_{2}}\left(  z\right)  $ and also
using semi-classical WKB approximation methods for more complicated cases, we
may estimate the behavior of $E_{n}\left(  z\right)  +E_{m_{1}m_{2}}\left(
z\right)  $ for small as well as large $z.$ This can then be used to discuss
the behavior of the quantum universe beyond the approximations described
above. In such attempts, with non-trivial $v\left(  \sigma\right)  ,v\left(
\alpha_{1},\alpha_{2}\right)  ,$ we find that $E_{n}\left(  z\right)
,E_{m_{1}m_{2}}\left(  z\right)  $ are generically not analytic near $z=0$  in
the complex $z$-plane (in the sense of cuts that extend to $z=0)$ so this
complicates the use of analyticity methods \cite{turok} to extract information
from these equations. We continue to investigate this approach and hope to
report more results in the future.

To conclude this section, an important remark is that unitarity is maintained
in the WdW treatment throughout gravity and antigravity, while the presence of
negative energy during antigravity is not of concern regarding fundamental
principles as already illustrated in the previous sections, especially with
the simpler computation based on Eq.(\ref{WdW}). 

\section{Negative energy in antigravity and observers in gravity}

To develop a physical understanding of negative kinetic energy we will discuss
several toy models that will include the analog of a background gravitational
field that switches sign between positive and negative kinetic energy. The
physical question is, what do observers in the gravity region detect about the
presence of a negative kinetic energy sector? Conceptually this is the analog
of a black box being probed by in/out signals detected at the exterior of the box.

In the field theory or particle examples discussed below a simple sign
function that is modeled after the \textquotedblleft antigravity loop
\textquotedblright\ in \cite{BST-antigravity} captures the main effect of
antigravity. This sign function is a simple device to answer questions that
arose repeatedly on unitarity and possible instability and is not necessarily
a solution to the gravitational field equations of some specific model.
Rather, it is used here only to capture the main effect of an antigravity
sector in a simple and solvable model. In the case of realistic applications
one would need to use a self consistent solution of matter and gravitational
equations (as in \cite{BST-antigravity}) as long as it captures the main
features of antigravity as in the simplified model background discussed here.

\subsection{Particle with time dependent kinetic energy flips
\label{particleFlip}}

A free particle with a relativistic (or non-relativistic) Hamiltonian that
switches sign as a function of time provides an example of a system
propagating in a background gravitational field that switches sign as in
Eq.(\ref{g-hat})
\begin{equation}
H=\varepsilon\left(  \left\vert t\right\vert -\frac{\Delta}{2}\right)
\times\sqrt{p^{2}+m^{2}}\;\text{ or \ }H=\varepsilon\left(  \left\vert
t\right\vert -\frac{\Delta}{2}\right)  \times\frac{p^{2}}{2m},
\label{Hparticle}%
\end{equation}
where $\varepsilon\left(  u\right)  \equiv$Sign$\left(  u\right)  .$ Such a
background captures some of the properties of the antigravity loop of
Bars-Steinhardt-Turok \cite{BST-antigravity}. The particle's phase space
$\left(  x,p\right)  $ can also represent more generally a typical generalized
degree of freedom in field theory or string field theory.

The momentum $p$ is conserved since $H$ is independent of $x,$ but the
velocity $\dot{x}=\partial H/\partial p=\varepsilon\left(  \left\vert
t\right\vert -\frac{\Delta}{2}\right)  \frac{p}{\sqrt{p^{2}+m^{2}}}$
alternates signs as shown below. The Hamiltonian is time dependent, so it is
not conserved.%
\[%
\begin{tabular}
[c]{|l|c|c|c|}\hline
$t:$ & $t<-\frac{\Delta}{2}$ & $-\frac{\Delta}{2}<t<\frac{\Delta}{2}\;\;$ &
$\;\;t>\frac{\Delta}{2}\;\;$\\\hline
$H_{\pm}:$ & $\sqrt{p^{2}+m^{2}}$ & $-\sqrt{p^{2}+m^{2}}$ & $\sqrt{p^{2}%
+m^{2}}$\\\hline
$x:$ & $\dot{x}=\frac{p}{\sqrt{p^{2}+m^{2}}}$ & $\dot{x}=-\frac{p}{\sqrt
{p^{2}+m^{2}}}$ & $\dot{x}=\frac{p}{\sqrt{p^{2}+m^{2}}}$\\\hline
\end{tabular}
\ \ \ \ \ \ \
\]
At the $t=\pm\Delta/2$ kinks the velocity vanishes if we define $\varepsilon
\left(  0\right)  =0.$ It is possible to make other models of what happens to
the velocity by replacing the sign function $\varepsilon\left(  z\right)  $ by
other time dependent kinky or smooth models; for example, if we replace
$\varepsilon\left(  z\right)  $ by $\left(  \varepsilon\left(  z\right)
\right)  ^{-1},$ then the velocity at the kinks changes sign at an infinite
value rather than at zero, while the momentum remains a constant in all
cases$.$

If the initial position before entering antigravity is $x_{i}\left(
t_{i}\right)  ,$ we compute the evolution at any time as follows (see Fig.1)
\begin{equation}%
\begin{array}
[c]{cc}%
t<-\frac{\Delta}{2}\;\;\;\;\;: & \;x\left(  t\right)  =x_{i}\left(
t_{i}\right)  +\frac{p}{\sqrt{p^{2}+m^{2}}}\left(  t-t_{i}\right) \\
-\frac{\Delta}{2}<t<\frac{\Delta}{2}: & \;x\left(  t\right)  =x\left(
-\frac{\Delta}{2}\right)  -\frac{p}{\sqrt{p^{2}+m^{2}}}\left(  t+\frac{\Delta
}{2}\right) \\
t>\frac{\Delta}{2}\;\;\;\;\;\;: & \;x\left(  t\right)  =x\left(  \frac{\Delta
}{2}\right)  +\frac{p}{\sqrt{p^{2}+m^{2}}}\left(  t-\frac{\Delta}{2}\right)
\end{array}
\label{method}%
\end{equation}
where $x\left(  -\frac{\Delta}{2}\right)  =x_{i}\left(  t_{i}\right)
+\frac{p}{\sqrt{p^{2}+m^{2}}}\left(  -\frac{\Delta}{2}-t_{i}\right)  ,$ and
$x\left(  \frac{\Delta}{2}\right)  =x_{i}\left(  t_{i}\right)  -\frac{p}%
{\sqrt{p^{2}+m^{2}}}\left(  3\Delta/2+t_{i}\right)  .$ The final position
$x_{f}\left(  t_{f}\right)  ,$ at a time $t_{f}$ after waiting long enough to
exit from antigravity, $t_{f}>\Delta/2,$ is%
\begin{equation}
x_{f}\left(  t_{f}\right)  =x_{i}\left(  t_{i}\right)  +\frac{p}{\sqrt
{p^{2}+m^{2}}}\left(  t_{f}-t_{i}-2\Delta\right)  . \label{delayedParticle}%
\end{equation}
The effect of antigravity during the interval, $-\frac{\Delta}{2}%
<t<\frac{\Delta}{2},$ is the backward excursion between the two kinks shown in
Fig.1 $.$ For observers waiting for the arrival of the particle at some
position $x_{f}\left(  t_{f}\right)  ,$ we see from Eq.(\ref{delayedParticle})
that antigravity causes a time delay by the amount of $2\Delta$ as compared to
the absence of antigravity. Hence there is a measurable signal, namely a time
delay, as an observable effect in comparing the presence and absence of antigravity.%

\begin{center}
\includegraphics[
height=1.7781in,
width=2.8665in
]%
{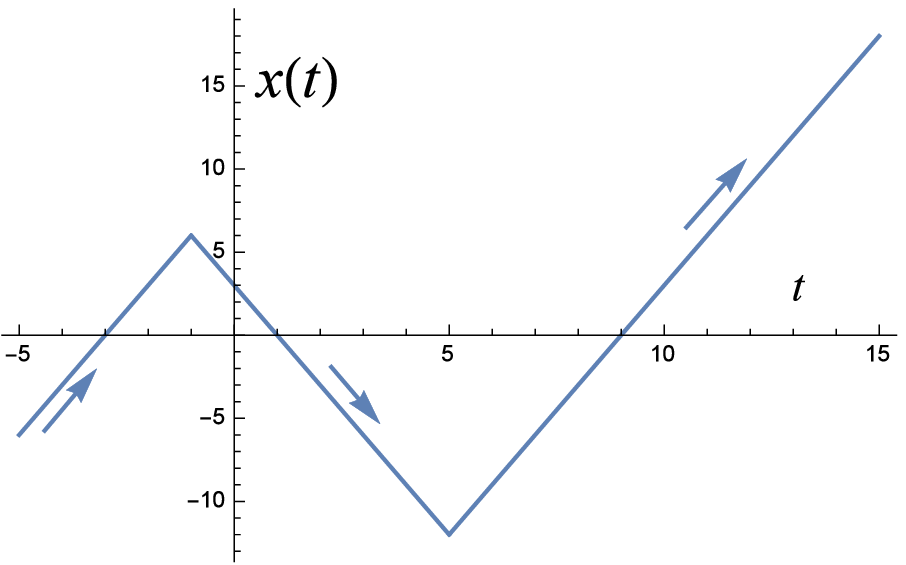}%
\\
Fig.1 - Propagation through antigravity.
\end{center}

A similar problem is analyzed at the quantum level by computing the transition
amplitude from an initial state $|x_{i},t_{i}\rangle$ to a final state
$|x_{f},t_{f}\rangle,$ requiring that the final observation is in the gravity
period, \textit{after} passing through the antigravity period. This is given
by
\begin{equation}%
\begin{array}
[c]{l}%
A_{fi}=\langle x_{f},t_{f}|e^{-\frac{i}{\hbar}H_{+}\left(  t_{f}-\frac{\Delta
}{2}\right)  }e^{-\frac{i}{\hbar}H_{-}\left(  \frac{\Delta}{2}-\frac{-\Delta
}{2}\right)  }e^{-\frac{i}{\hbar}H_{+}\left(  \frac{-\Delta}{2}-t_{i}\right)
}|x_{i},t_{i}\rangle\\
=\langle x_{f},t_{f}|e^{-\frac{i}{\hbar}H\left(  t_{f}-t_{i}-2\Delta\right)
}|x_{i},t_{i}\rangle\\
=\sqrt{\frac{m}{2\pi i\hbar\left(  t_{f}-t_{i}-2\Delta\right)  }}\exp\left(
\frac{i~m~\left(  x_{f}-x_{i}-2\Delta\right)  ^{2}}{2\hbar\left(  t_{f}%
-t_{i}-2\Delta\right)  }\right)
\end{array}
\end{equation}
The last expression is for the case of a non-relativistic particle with
$H_{\pm}=\pm H=\pm p^{2}/2m.$ The exponentials involving $H_{\pm}$ are
simplified because $H_{\pm}$ commute with each other, allowing the combination
of the exponentials into a single exponential. Thus, the effect of the
intermediate antigravity period is to cause only a time delay just as in the
classical solution above. Note also that there are no unitarity problems; the
evolution operator is unitary, and norms of states are positive, at all stages.

\subsection{Particle with space dependent kinetic energy flips}

Consider a non-relativistic particle with a total energy Hamiltonian that
switches sign in different regions of space, for example
\begin{equation}
H=\varepsilon\left(  \left\vert x\right\vert -\frac{\Delta}{2}\right)
\times\left(  \frac{p^{2}}{2m}+V\left(  x\right)  \right)  .
\end{equation}
This is another example of a system propagating in a background gravitational
field that switches sign as in Eq.(\ref{g-hat}). In this case energy is
conserved since there is no explicit time dependence in $H$. Therefore at
generic energies, $E=\left(  \frac{p^{2}}{2m}+V\left(  x\right)  \right)  $,
the particle cannot cross the boundaries at $\left\vert x\right\vert
=\frac{\Delta}{2}$ since the Hamiltonian would flip sign and this would
contradict the energy conservation. Hence if the particle is in the the
gravity region, $\left\vert x\right\vert >\frac{\Delta}{2},$ it stays there,
and if it is in the antigravity region, $\left\vert x\right\vert <\frac
{\Delta}{2},$ it stays there. However, the particle can cross from gravity to
antigravity and back again to gravity at zero energy $\frac{p^{2}}%
{2m}+V\left(  x\right)  =0.$ This is similar to the geodesics in a black hole
that cross from gravity to antigravity \cite{ABJ-blackhole}.

\subsection{Free massless scalar field with sign flipping kinetic energy
\label{NoMass}}

Consider a free massless scalar field in flat space with a time dependent
background field that causes sign flips of the kinetic energy as a function of
time%
\begin{equation}
S=-\frac{1}{2}\int d^{4}x\;\varepsilon\left(  \left\vert x^{0}\right\vert
-\frac{\Delta}{2}\right)  \;\partial^{\mu}\phi\left(  x\right)  \partial_{\mu
}\phi\left(  x\right)  .
\end{equation}
The factor $\varepsilon\left(  \left\vert x^{0}\right\vert -\frac{\Delta}%
{2}\right)  $ can be viewed as a gravitational background field of the form,
$\sqrt{-g}g^{\mu\nu}\partial_{\mu}\phi\partial_{\nu}\phi,$ with $g^{\mu\nu
}\left(  x\right)  =\varepsilon\left(  \left\vert x^{0}\right\vert
-\frac{\Delta}{2}\right)  \eta^{\mu\nu}$ and $\sqrt{-g}=1.$ This sign flipping
metric should be regarded as an example of a geometry that spans the union of
the gravity and antigravity regions, as in Eq.(\ref{g-hat}). We proceed to
analyze the time evolution of this system. Let the on-shell initial field
configuration at time $x_{i}^{0}<\left(  -\Delta/2\right)  $ be defined by
\begin{equation}
\phi_{i}\left(  \vec{x}_{i},x_{i}^{0}\right)  =\int\frac{d^{3}p}{\left(
2\pi\right)  ^{3/2}2\left\vert p\right\vert }\left(  a\left(  \vec{p}\right)
e^{-i\left\vert p\right\vert x_{i}^{0}+i\vec{p}\cdot\vec{x}_{i}}+\bar
{a}\left(  \vec{p}\right)  e^{i\left\vert p\right\vert x_{i}^{0}-i\vec{p}%
\cdot\vec{x}_{i}}\right)
\end{equation}
The general solution for $\phi\left(  \vec{x},x^{0}\right)  $ evolved up to a
final time $x_{f}^{0}>\Delta/2$ is then given by (using the method in
Eq.(\ref{method}))%
\begin{equation}
\phi_{f}\left(  \vec{x}_{f},x_{f}^{0}\right)  =\int\frac{d^{3}p}{\left(
2\pi\right)  ^{3/2}2\left\vert p\right\vert }\left(
\begin{array}
[c]{c}%
a\left(  \vec{p}\right)  e^{-i\left\vert p\right\vert \left(  x_{f}^{0}%
-x_{i}^{0}-2\Delta\right)  +i\vec{p}\cdot\vec{x}_{f}}\\
+\bar{a}\left(  \vec{p}\right)  e^{i\left\vert p\right\vert \left(  x_{f}%
^{0}-x_{i}^{0}-2\Delta\right)  -i\vec{p}\cdot\vec{x}_{f}}%
\end{array}
\right)  .
\end{equation}
This shows that for initial/final observations, that are strictly outside of
the antigravity period, the effect of the antigravity period is only a time
delay as compared to the complete absence of antigravity. The time evolution
of the field in the interim period is just like the time evolution of the
particle as shown in Fig.1. For more details on the classical evolution of the
field in the interim period see the case of the massive field in section
(\ref{Mass}), and take the zero mass limit.

An important remark is that the multiparticle Hilbert space $\left\{  |\vec
{p}_{1},\vec{p}_{2}\cdots\vec{p}_{n}\rangle\right\}  $ is the Fock space
constructed from the creation operators applied on the vacuum defined by
$a\left(  \vec{p}\right)  |0\rangle=0,$ namely $|\vec{p}_{1},\vec{p}_{2}%
\cdots\vec{p}_{n}\rangle\equiv\bar{a}\left(  \vec{p}_{1}\right)  \bar
{a}\left(  \vec{p}_{2}\right)  \cdots\bar{a}\left(  \vec{p}_{n}\right)
|0\rangle.$ This time independent Fock space is the complete Hilbert space
that can be used during gravity or antigravity. It is clearly unitary since it
is the same Hilbert space that is independent of the existence of an
antigravity period (i.e. same as the $\Delta=0$ case). This shows that there
is no unitarity problem due to the presence of the antigravity period.

However, there is negative kinetic energy during antigravity, seen as follows.
The time dependent Hamiltonian for this system is
\begin{equation}
H\left(  x^{0}\right)  =\left\{
\begin{array}
[c]{l}%
+H,\;\text{for }t<-\frac{\Delta}{2}\\
-H,\;\text{for }-\frac{\Delta}{2}<t<\frac{\Delta}{2}\\
+H,\;\text{for }t>\frac{\Delta}{2}%
\end{array}
\right.
\end{equation}
where $H,$ which is constructed from the quantum creation-annihilation
operators as usual, is time independent. So there seems to be a possible
source of instability due to negative energy during antigravity. For freely
propagating particles there are no transitions that alter the energy, so no
questions arise, it is only when there are interactions that an effect may be
observed due to transitions created by the negative energy sector. The effect
of interactions, as observed by detectors in the gravity sector, is analogous
to the case of a time dependent Hamiltonian as discussed in simple examples
below in section (\ref{interact}). Hence, the presence of a sector with
negative kinetic energy is not a fundamental problem in the quantum theory.

Nevertheless, the antigravity sector, with or without interactions, is the
source of interesting physical signals for the observers in the gravity
sectors. For example, in the absence of additional interactions, consider the
quantum propagator that corresponds to initial/final states in the two gravity
sectors $\left\vert x^{0}\right\vert >\Delta/2$. The transition amplitude from
an initial state in gravity $\left(  x_{i}^{0}<-\Delta/2\right)  $ to a final
state in gravity $\left(  x_{f}^{0}>\Delta/2\right)  $, after the field
evolves through antigravity, is given by%
\begin{align*}
A_{fi}  &  =\langle\phi_{f}\left(  x_{f}\right)  |e^{-\frac{i}{\hbar}%
H_{+}\left(  t_{f}-\frac{\Delta}{2}\right)  }e^{-\frac{i}{\hbar}H_{-}\left(
\frac{\Delta}{2}-\frac{-\Delta}{2}\right)  }e^{-\frac{i}{\hbar}H_{+}\left(
\frac{-\Delta}{2}-t_{i}\right)  }|\phi_{i}\left(  x_{i}\right)  \rangle\\
&  =\langle\phi_{f}\left(  \vec{x}_{f},x_{f}^{0}\right)  |e^{-\frac{i}{\hbar
}H\left(  t_{f}-t_{i}-2\Delta\right)  }|\phi_{i}\left(  \vec{x}_{i},x_{i}%
^{0}\right)  \rangle
\end{align*}
Here, $|\phi\left(  x\right)  \rangle$ is defined as the 1-particle state in
the quantum theory which is created by applying the quantum field $\hat{\phi
}\left(  x\right)  $ on the oscillator vacuum $a\left(  \vec{p}\right)
|0\rangle=0,$
\begin{equation}
|\phi\left(  \vec{x},x^{0}\right)  \rangle=\hat{\phi}\left(  x\right)
|0\rangle=\int d^{3}p\frac{e^{i\left\vert p\right\vert x^{0}-i\vec{p}\cdot
\vec{x}}}{\left(  2\pi\right)  ^{3/2}2\left\vert p\right\vert }\bar{a}\left(
\vec{p}\right)  |0\rangle.
\end{equation}
Then we obtain
\begin{equation}
A_{fi}=\int d^{3}p\frac{e^{i\left\vert p\right\vert \left(  x_{f}^{0}%
-x_{i}^{0}-2\Delta\right)  -i\vec{p}\cdot\left(  \vec{x}_{f}-\vec{x}%
_{i}\right)  }}{\left(  2\pi\right)  ^{3/2}2\left\vert p\right\vert }.
\end{equation}
This is the propagator for a free massless particle. From this expression it
is clear that the effect of antigravity on the result for the transition
amplitude $A_{fi}$ is only a time delay by an amount of $2\Delta$ as compared
to the same quantity in the complete absence of antigravity$.$ The same
general statement holds true for the transition amplitudes for multi-particle
states. Clearly there is no particle production due to antigravity in the case
of free massless particles. This will be contrasted with the case of massive
particles in section (\ref{Mass}).

Of course, if there are field interactions, there will be additional effects,
but none of those are \`{a} priori problematic from the point of view of
fundamental principles.

\subsection{Particle with flipping kinetic energy while interacting in a
potential \label{interact}}

To learn more about the effects of antigravity we now add an interaction term
that does not flip sign during antigravity. We first investigate the case of a
single degree of freedom whose kinetic energy flips sign during antigravity.
This phase space $\left(  x,p\right)  $ should be thought of as a generalized
coordinate associated with any single degree of freedom within local field
theory or string field theory (after integrating out all other degrees of
freedom), but in its simplest form it can be regarded as representing a
particle moving in one dimension.

We discuss a simple model described by the Hamiltonian%

\begin{equation}
H=\varepsilon\left(  \left\vert t\right\vert -\frac{\Delta}{2}\right)
\times\frac{p^{2}}{2m}+\frac{m\omega^{2}}{2}x^{2}.
\end{equation}
This is a time dependent Hamiltonian that has two different forms, $H_{\pm},$
during different periods of time as follows%
\begin{equation}%
\begin{tabular}
[c]{|l|l|l|l|}\hline
$t:$ & $t<-\frac{\Delta}{2}$ & $-\frac{\Delta}{2}<t<\frac{\Delta}{2}\;\;$ &
$\;\;t>\frac{\Delta}{2}\;\;$\\\hline
$H_{\pm}:$ & $\left(  \frac{p^{2}}{2m}+\frac{m\omega^{2}x^{2}}{2}\right)  $ &
$\left(  -\frac{p^{2}}{2m}+\frac{m\omega^{2}x^{2}}{2}\right)  $ & $\left(
\frac{p^{2}}{2m}+\frac{m\omega^{2}x^{2}}{2}\right)  $\\\hline
\end{tabular}
\end{equation}
During gravity, $\left\vert t\right\vert >\frac{\Delta}{2},$ the Hamiltonian
$H_{+}$ is the familiar harmonic oscillator Hamiltonian with a well defined
quantum state, so all energies are positive. But during antigravity,
$-\frac{\Delta}{2}<t<\frac{\Delta}{2},$ the Hamiltonian $H_{-}$ has no bottom,
so all positive and negative energies are permitted. Does this pose an
instability problem for the entire system? The answer is that, as in the
simpler cases already illustrated above, there is no such problem from the
perspective of observers in gravity.

A complete basis for a unitary Hilbert space may be defined to be the positive
norm complete Fock space associated with the usual harmonic oscillator
Hamiltonian $H_{+}$ whose energy eigenvalues are strictly positive. The
eigenstates of $H_{-}$ are also positive norm and define another complete
unitary basis. Clearly one complete basis may be expanded in terms of another
complete basis, so the usual Fock space basis is sufficient to analyze the
complete system, including its evolution through antigravity. This shows that
the interacting problem that includes antigravity is an ordinary time
dependent problem in quantum mechanics. There are no unitarity problems, and
the presence of antigravity is analyzed below as a regular problem of a time
dependent Hamiltonian, without encountering any fundamental problems of principles.

A technical remark may be useful: this model can be treated group
theoretically by using the properties of SL$(2,R)$ representations. Note that
the three Hermitian quantum operators $\left(  x^{2},p^{2},\frac{1}{2}\left(
xp+px\right)  \right)  $ form the algebra of SL$\left(  2,R\right)  $ under
quantum commutation rules $\left[  x,p\right]  =i\hbar$. The Hamiltonian
$H_{+}$ is proportional to the compact generator $J_{0}$ of SL$\left(
2,R\right)  ,$ while $H_{-}$ is proportional to one of the non-compact
generators $J_{1}$. The second non-compact generator $J_{2}$ appears in the
commutator $\left[  H_{+},H_{-}\right]  .$ Explicitly,
\begin{equation}
J_{0}\equiv\frac{1}{2\hbar\omega}H_{+},\;J_{1}\equiv\frac{1}{2\hbar\omega
}H_{-},\;J_{2}=\frac{1}{4\hbar}\left(  xp+px\right)  .
\end{equation}
The $\left(  J_{0},J_{1},J_{2}\right)  $ form the standard Lie algebra of
SL$\left(  2,R\right)  .$ Since these $J_{0,1,,2}$ are Hermitian operators,
the corresponding quantum states which are labelled as $|j,\mu\rangle$ form a
unitary representation of SL$\left(  2,R\right)  .$ The quantum number $\mu$
is associated with the eigenvalues of $J_{0}$ (which is basically the
eigenvalues of $H_{+})$ while $j\left(  j+1\right)  $ is associated with the
eigenvalues of Casimir operator $C_{2}$ that commutes with all the generators,
$C_{2}\equiv J_{0}^{2}-J_{1}^{2}-J_{2}^{2}.$ For the present construction,
keeping track of the orders of operators $\left(  x,p\right)  $ one finds that
$C_{2}$ is a constant, $C_{2}=-3/16=j\left(  j+1\right)  ,$ which yields two
solutions $j=-\frac{3}{4}$ or $j=-\frac{1}{4}.$ Hence the spectrum of this
theory, including the properties of $H_{\pm}$ can be thought of consisting of
two infinite dimensional irreducible unitary representations of SL$\left(
2,R\right)  .$ For $j=-\frac{3}{4}$ or $j=-\frac{1}{4}$ these are positive
discrete series representations. The allowed values of $\mu$ are given by
$\mu=j+1+k$ where $k=0,1,2,\cdots$ is an integer. We see that the two
representations taken together correspond to the spectrum of $H_{+}$, which is
the spectrum of the harmonic oscillator given by $E_{n}=\omega\left(
n+\frac{1}{2}\right)  \Leftrightarrow2\omega\mu,$ with even $n=2k$
corresponding to $j=-3/4$ and odd $n=2k+1$ corresponding to $j=-1/4.$ Hence
the basis $|j,\mu\rangle$ form a complete set of eigenstates for the observers
in the gravity sector of the theory.

How about the antigravity sector? Since the corresponding Hamiltonian is
$H_{-},$ a complete set of eigenstates corresponds to diagonalizing the
non-compact generator $J_{1}$ instead of the compact generator $J_{0}.$ Either
way the Casimir operator is the same; hence diagonalizing $J_{1}\rightarrow q$
provides another unitary basis $|j,q\rangle$ for the same unitary
representations of SL$\left(  2,R\right)  .$ The spectrum of $J_{1},$
$J_{1}|j,q\rangle=q$ $|j,q\rangle,$ is continuous $q$ on the real line since
this is a non-compact generator of SL$\left(  2,R\right)  .$ This antigravity
basis is also a complete unitary basis for this Hamiltonian that includes both
sectors $H_{\pm}$. One basis can be expanded in terms of the other,
$|j,q\rangle=\sum_{\mu=j+1}^{\infty}|j,\mu\rangle\langle j,\mu|j,q\rangle,$
where the expansion coefficients $\langle j,\mu|j,q\rangle=U_{\mu,q}^{\left(
j\right)  }$ is a unitary transformation for each $j=-\frac{3}{4}$ or
$-\frac{1}{4}$.

Therefore it doesn't matter which basis we use to analyze the quantum
properties of this Hamiltonian. Using the discrete basis $|j,\mu\rangle$ which
is more convenient to analyze the physics in the gravity sector, in no way
excludes the antigravity sector from making its effects felt for observers in
the gravity sector.

With this understanding of this simple quantum system, we now analyze the
transition amplitudes $A_{fi}$ for an initial state $|i\rangle$ to propagate
to a final state both in the gravity sector. We define $|i\rangle,|f\rangle$
at the two edges of the antigravity sector, at times $t_{i}=-\Delta/2$ and
$t_{f}=\Delta/2$. Moving $t_{i},t_{f}$ to other arbitrary times in the gravity
sector is trivial since we can write $|i\rangle=e^{-iH_{+}\left(
-\Delta/2-t_{i}\right)  }|i,t_{i}\rangle$ and $|f\rangle=e^{-iH_{+}\left(
t_{f}-\Delta/2\right)  }|f,t_{f}\rangle$ and we know how $H_{+}$ acts on any
linear combination of harmonic oscillator states $|i\rangle,|f\rangle$. Hence
we have
\begin{equation}
A_{fi}=\langle f|e^{-\frac{i}{\hbar}\Delta H_{-}}|i\rangle
\end{equation}
where $|i\rangle,|f\rangle$ are arbitrary states in the gravity sector. If we
take any two states in the SL$\left(  2,R\right)  $ basis $|j,\mu\rangle$,
this becomes%
\begin{equation}
A_{fi}=\langle j,\mu_{f}|e^{-i\frac{\Delta}{2\omega}J_{1}}|j,\mu_{i}\rangle.
\end{equation}
This is just the matrix representation of a group element of SL$\left(
2,R\right)  $ in a unitary representation labelled by $j=-\frac{3}{4}$ or
$-\frac{1}{4}.$ It must be the same $j$ for both the initial and final states,
i.e. there is a selection rule because there can be no transitions at all from
$j=-\frac{3}{4}$ to $j=-\frac{1}{4}$ and vice-versa.

This quantity can be computed by using purely group theoretical means, but it
is perhaps more instructive to use the standard harmonic oscillator
creation/annihilation operators to evaluate it. Then we can write
\begin{equation}
H_{+}=\hbar\omega\left(  a^{\dagger}a+\frac{1}{2}\right)  ,\;\;H_{-}%
=\frac{\hbar\omega}{2}\left(  a^{\dagger2}+a^{2}\right)  .
\end{equation}
We have used this form to compute the transition amplitude
\begin{equation}
A_{fi}=\langle f|e^{-\frac{i}{\hbar}\Delta H_{-}}|i\rangle=\langle
f|e^{-i\frac{\omega\Delta}{2}\left(  a^{\dagger2}+a^{2}\right)  }|i\rangle,
\end{equation}
by taking initial/final states to be the number states or the coherent states
of the harmonic oscillator. To perform the computation we use the following
identity%
\begin{equation}
e^{-i\frac{\omega\Delta}{2}\left(  a^{\dagger2}+a^{2}\right)  }=e^{-\frac
{i}{2}\tanh\left(  \omega\Delta\right)  a^{\dagger2}}\left(  \cosh\left(
\omega\Delta\right)  \right)  ^{-\left(  a^{\dagger}a+\frac{1}{2}\right)
}e^{-\frac{i}{2}\tanh\left(  \omega\Delta\right)  a^{2}}.
\end{equation}
For initial/final coherent states, $|z_{i}\rangle$ \& $|z_{f}\rangle$ for
observers in gravity, we define the transition amplitude for normalized states
as, $A\left(  z_{f},z_{i}\right)  =\langle z_{f}|e^{-\frac{i}{\hbar}\Delta
H_{-}}|z_{i}\rangle/\sqrt{\langle z_{f}|z_{f}\rangle\langle z_{i}|z_{i}%
\rangle},$ which yields
\begin{equation}
\left\vert A\left(  z_{f},z_{i}\right)  \right\vert ^{2}=\frac{e^{-\left\vert
z_{f}\right\vert ^{2}-\left\vert z_{i}\right\vert ^{2}+\frac
{2\operatorname{Re}\left(  z_{i}\bar{z}_{f}\right)  }{\cosh\left(
\omega\Delta\right)  }}e^{\tanh\left(  \omega\Delta\right)  \operatorname{Im}%
\left(  \bar{z}_{f}^{2}e^{-i\omega\Delta}+z_{i}^{2}e^{i\omega\Delta}\right)
}}{\cosh\left(  \omega\Delta\right)  }.
\end{equation}
This should be compared to the absence of antigravity when $\Delta=0$, namely
$\left\vert A\left(  z_{f},z_{i}\right)  \right\vert ^{2}\overset
{\Delta\rightarrow0}{=}e^{-\left\vert z_{f}-z_{i}\right\vert ^{2}}.$

Similarly, for initial/final number eigenstates $|n\rangle$ \& $|m\rangle$ of
the Hamiltonian $H_{+}=\left(  \frac{p^{2}}{2m}+\frac{m\omega^{2}x^{2}}%
{2}\right)  =\hbar\omega\left(  a^{\dagger}a+\frac{1}{2}\right)  \;$for
observers in gravity, we obtain
\begin{equation}
A_{mn}=\sqrt{\frac{m!n!e^{i\omega\Delta(n+m+1)}}{\left(  \cosh\left(
\omega\Delta\right)  \right)  ^{m+n+1}}}\sum_{k=0}^{\min\left(  m,n\right)
}\frac{\left(  \frac{1}{2i}\sinh\left(  \omega\Delta\right)  \right)
^{\frac{m+n}{2}-k}}{k!\left(  \frac{m-k}{2}\right)  !\left(  \frac{n-k}%
{2}\right)  !}\text{,}%
\end{equation}
where $\left(  m,n,k\right)  $ are all even or all odd. This gives
\[
\left\vert A_{mn}\right\vert ^{2}=\left(
\begin{array}
[c]{c}%
\left(  _{2}F_{1}(-\left[  \frac{m}{2}\right]  ,-\left[  \frac{n}{2}\right]
;\left(  1-\frac{\left(  -1\right)  ^{m}}{2}\right)  ;\frac{-1}{\sinh
^{2}\left(  \omega\Delta\right)  })\right)  ^{2}\\
\times\frac{\left(  m!\right)  \left(  n!\right)  \left(  \frac{1}{2}%
\tanh\left(  \omega\Delta\right)  \right)  ^{2\left(  \left[  \frac{m}%
{2}\right]  +\left[  \frac{n}{2}\right]  \right)  }}{\left(  \left[  \frac
{m}{2}\right]  !\left[  \frac{n}{2}\right]  !\right)  ^{2}~\left(
\cosh\left(  \omega\Delta\right)  \right)  ^{2-\left(  -1\right)  ^{m}}}%
\end{array}
\right)
\]
where $_{2}F_{1}\left(  a,b;c;z\right)  $ is the hypergeometric function,
$\left[  \frac{m}{2}\right]  $ means the integer part of $m/2$, and $\left(
m,n\right)  $ are both even or both odd. Special cases are%
\begin{equation}%
\begin{array}
[c]{c}%
\left\vert A_{00}\right\vert ^{2}=\frac{1}{\cosh\left(  \omega\Delta\right)
},\ \;\left\vert A_{2M,0}\right\vert ^{2}=\frac{\left(  2M\right)  !}%
{2^{2M}\left(  M!\right)  ^{2}}\frac{\left(  \tanh\left(  \omega\Delta\right)
\right)  ^{2M}}{\cosh\left(  \omega\Delta\right)  },\\
\left\vert A_{11}\right\vert ^{2}=\frac{1}{\cosh^{3}\left(  \omega
\Delta\right)  },\;A_{2M+1,1}=\frac{\left(  2M+1\right)  !}{2^{2M}\left(
M!\right)  ^{2}}\frac{\left(  \tanh\left(  \omega\Delta\right)  \right)
^{2M}}{\cosh^{3}\left(  \omega\Delta\right)  }.
\end{array}
\end{equation}
As compared to the absence of antigravity, $\Delta=0,$ when there are no
transitions, we see that antigravity causes an observable effect. Clearly,
these transition amplitudes are well behaved, and do not blow up for large
$\Delta.$ Unitarity is obeyed: one may verify explicitly that the sum over all
states is 100\% probability, $\sum_{m}\left\vert A_{mn}\right\vert ^{2}=1,$
for all fixed $n,$ and similarly $\int\frac{d^{2}z_{f}}{\pi}~\left\vert
A\left(  z_{f},z_{i}\right)  \right\vert ^{2}=1$ for all fixed $z_{i}.$

\subsection{Massive scalar field with sign flipping kinetic energy
\label{Mass}}

This system has some similarities to the interacting particle above but it is
not quite the same. The action is%
\begin{equation}
S=\frac{1}{2}\int d^{d}x\;\left[  -\varepsilon\left(  \left\vert
x^{0}\right\vert -\frac{\Delta}{2}\right)  \;\partial^{\mu}\phi\left(
x\right)  \partial_{\mu}\phi\left(  x\right)  -m^{2}\phi^{2}\left(  x\right)
\right]
\end{equation}
As in the case of the massless field in section (\ref{NoMass}), the factor
$\varepsilon\left(  \left\vert x^{0}\right\vert -\frac{\Delta}{2}\right)  $
can be viewed as a gravitational background field of the form, $\sqrt
{-g}g^{\mu\nu}\partial_{\mu}\phi\partial_{\nu}\phi,$ with $g^{\mu\nu}\left(
x\right)  =\varepsilon\left(  \left\vert x^{0}\right\vert -\frac{\Delta}%
{2}\right)  \eta^{\mu\nu}$ and $\sqrt{-g}=1,$ that spans the union of the
gravity and antigravity regions, as explained in Eq.(\ref{g-hat}). The mass
term does not flip sign. Note that, due to the non-zero mass, this is
\textit{not} a Weyl invariant action, but we will investigate it anyway to
learn about the properties of such a system.

In momentum space, using the notation $x^{0}=t$, we have%
\begin{equation}
\phi\left(  \vec{x},t\right)  =\int\frac{d^{d-1}p}{\left(  2\pi\right)
^{\left(  d-1\right)  /2}}\phi_{p}\left(  t\right)  e^{i\vec{p}\cdot\vec{x}}%
\end{equation}
We rewrite the action in momentum space as%
\begin{equation}
S=\frac{1}{2}\int dt\int d^{d-1}p\;\left[
\begin{array}
[c]{c}%
\varepsilon\left(  \left\vert t\right\vert -\frac{\Delta}{2}\right)  \left[
\begin{array}
[c]{c}%
\dot{\phi}_{p}\left(  t\right)  \dot{\phi}_{-p}\left(  t\right) \\
-\vec{p}^{2}\phi_{p}\left(  t\right)  \phi_{-p}\left(  t\right)
\end{array}
\right] \\
-m^{2}\phi_{p}\left(  t\right)  \phi_{-p}\left(  t\right)
\end{array}
\right]
\end{equation}
The equation of motion is
\begin{equation}
\partial_{t}\left(  \varepsilon\left(  \left\vert t\right\vert -\frac{\Delta
}{2}\right)  \partial_{t}\phi_{p}\left(  t\right)  \right)  +\left[
\varepsilon\left(  \left\vert t\right\vert -\frac{\Delta}{2}\right)  \vec
{p}^{2}+m^{2}\right]  \phi_{p}\left(  t\right)  =0
\end{equation}
The solutions in separate regions of time are (similar to (\ref{method}))%
\begin{equation}%
\begin{array}
[c]{c}%
t<-\frac{\Delta}{2}:\;\phi_{p}^{A}\left(  t\right)  =\left(
\begin{array}
[c]{c}%
A_{p}^{+}e^{-i\sqrt{\vec{p}^{2}+m^{2}}\left(  t+\frac{\Delta}{2}\right)  }\\
+A_{p}^{-}~e^{i\sqrt{\vec{p}^{2}+m^{2}}\left(  t+\frac{\Delta}{2}\right)  }%
\end{array}
\right) \\
-\frac{\Delta}{2}<t<\frac{\Delta}{2}:\phi_{p}^{B}\left(  t\right)  =\left(
\begin{array}
[c]{c}%
B_{p}^{+}~e^{-i\sqrt{\vec{p}^{2}-m^{2}}\left(  t+\frac{\Delta}{2}\right)  }\\
+B_{p}^{-}e^{i\sqrt{\vec{p}^{2}-m^{2}}\left(  t+\frac{\Delta}{2}\right)  }%
\end{array}
\right) \\
t>\frac{\Delta}{2}:\;\phi_{p}^{C}\left(  t\right)  =\left(
\begin{array}
[c]{c}%
C_{p}^{+}~e^{-i\sqrt{\vec{p}^{2}+m^{2}}\left(  t-\frac{\Delta}{2}\right)  }\\
+C_{p}^{-}e^{i\sqrt{\vec{p}^{2}+m^{2}}\left(  t-\frac{\Delta}{2}\right)  }%
\end{array}
\right)
\end{array}
\end{equation}
We need to match the field $\phi_{p}\left(  t\right)  $ and its canonical
momentum, $\varepsilon\left(  \left\vert t\right\vert -\frac{\Delta}%
{2}\right)  \partial_{t}\phi_{p}\left(  t\right)  ,$ at each boundary
$t=\pm\Delta/2$%
\begin{equation}%
\begin{array}
[c]{c}%
\phi_{p}^{A}\left(  -\frac{\Delta}{2}\right)  =\phi_{p}^{B}\left(
-\frac{\Delta}{2}\right)  ,\text{ and }\dot{\phi}_{p}^{A}\left(  -\frac
{\Delta}{2}\right)  =-\dot{\phi}_{p}^{B}\left(  -\frac{\Delta}{2}\right)  ,\\
\phi_{p}^{C}\left(  +\frac{\Delta}{2}\right)  =\phi_{p}^{B}\left(
+\frac{\Delta}{2}\right)  ,\text{ and }\dot{\phi}_{p}^{C}\left(  +\frac
{\Delta}{2}\right)  =-\dot{\phi}_{p}^{B}\left(  +\frac{\Delta}{2}\right)  ,
\end{array}
\end{equation}
Note the sign flip of $\dot{\phi}$ at $t=\pm\Delta/2$ although the canonical
momentum does not flip. This gives four equations to relate $C_{p}^{\pm}$ and
$B_{p}^{\pm}$ to $A_{p}^{\pm}$ as follows%
\begin{equation}%
\begin{array}
[c]{l}%
A_{p}^{+}+A_{p}^{-}=B_{p}^{+}+B_{p}^{-}\\
A_{p}^{+}-A_{p}^{-}=-\left(  B_{p}^{+}~-B_{p}^{-}\right)  \frac{\sqrt{\vec
{p}^{2}-m^{2}}}{\sqrt{\vec{p}^{2}+m^{2}}}\\
C_{p}^{+}+C_{p}^{-}=B_{p}^{+}e^{-i\sqrt{\vec{p}^{2}-m^{2}}\Delta}+B_{p}%
^{-}e^{i\sqrt{\vec{p}^{2}-m^{2}}\Delta}\\
C_{p}^{+}-C_{p}^{-}=-\left(
\begin{array}
[c]{c}%
B_{p}^{+}e^{-i\sqrt{\vec{p}^{2}-m^{2}}\Delta}\\
-B_{p}^{-}e^{i\sqrt{\vec{p}^{2}-m^{2}}\Delta}%
\end{array}
\right)  \frac{\sqrt{\vec{p}^{2}-m^{2}}}{\sqrt{\vec{p}^{2}+m^{2}}}%
\end{array}
\end{equation}
The solution determines $B_{p}^{\pm}$ and $C_{p}^{\pm}$ in terms of
$A_{p}^{\pm},$
\begin{align}
\left(
\begin{array}
[c]{c}%
C_{p}^{+}\\
C_{p}^{-}%
\end{array}
\right)   &  =\left(
\begin{array}
[c]{cc}%
\alpha & \beta^{\ast}\\
\beta & \alpha^{\ast}%
\end{array}
\right)  \left(
\begin{array}
[c]{c}%
A_{p}^{+}\\
A_{p}^{-}%
\end{array}
\right) \label{transBeta}\\
\left(
\begin{array}
[c]{c}%
B_{p}^{+}\\
B_{p}^{-}%
\end{array}
\right)   &  =\left(
\begin{array}
[c]{cc}%
\frac{1}{2}-\frac{\sqrt{\vec{p}^{2}+m^{2}}}{2\sqrt{\vec{p}^{2}-m^{2}}} &
\;\;\frac{1}{2}+\frac{\sqrt{\vec{p}^{2}+m^{2}}}{2\sqrt{\vec{p}^{2}-m^{2}}}\\
\frac{1}{2}+\frac{\sqrt{\vec{p}^{2}+m^{2}}}{2\sqrt{\vec{p}^{2}-m^{2}}} &
\;\;\frac{1}{2}-\frac{\sqrt{\vec{p}^{2}+m^{2}}}{2\sqrt{\vec{p}^{2}-m^{2}}}%
\end{array}
\right)  \left(
\begin{array}
[c]{c}%
A_{p}^{+}\\
A_{p}^{-}%
\end{array}
\right)
\end{align}
where $\left(  \alpha,\beta\right)  $ are the parameters of a Bogoliubov
transformation (an SU$\left(  1,1\right)  $ group transformation)%
\begin{equation}%
\begin{array}
[c]{l}%
\alpha=\cos\left(  \Delta\sqrt{\vec{p}^{2}-m^{2}}\right)  +i\frac{\vec{p}%
^{2}\sin\left(  \Delta\sqrt{\vec{p}^{2}-m^{2}}\right)  }{\sqrt{\left(  \vec
{p}^{2}\right)  ^{2}-m^{4}}},\\
\beta=i\frac{m^{2}\sin\left(  \Delta\sqrt{\vec{p}^{2}-m^{2}}\right)  }%
{\sqrt{\left(  p^{2}\right)  ^{2}-m^{4}}},\\
\left\vert \alpha\right\vert ^{2}-\left\vert \beta\right\vert ^{2}=1.
\end{array}
\end{equation}
Assume the incoming state $\phi_{p}^{A}\left(  t\right)  $ has only positive
frequency, meaning $A_{p}^{-}=0.$ Then we see that (unlike the massless case
in section (\ref{NoMass})) negative frequency fluctuations are produced in the
final state $\phi_{p}^{C}\left(  t\right)  $ since according to
Eq.(\ref{transBeta}), $C_{p}^{-}=\beta A_{p}^{+}.$ The corresponding
probability amplitude for particle production is
\begin{equation}
\left(  C_{p}^{-}/A_{p}^{+}\right)  =\beta=i\frac{\sin\left(  m\Delta
\sqrt{\vec{p}^{2}/m^{2}-1}\right)  }{\sqrt{\left(  \vec{p}^{2}/m^{2}\right)
^{2}-1}}.
\end{equation}
The produced particle number density (particles per unit volume) is the
integral of $\left\vert \beta\right\vert ^{2}$ over all momenta%
\begin{equation}%
\begin{array}
[c]{c}%
n\left(  m,\Delta\right)  =\int d^{d-1}p\left\vert \beta\right\vert ^{2}=\int
d^{d-1}p\frac{\sin^{2}\left(  m\Delta\sqrt{\left(  \vec{p}^{2}/m^{2}\right)
-1}\right)  }{\left\vert \left(  \vec{p}^{2}/m^{2}\right)  ^{2}-1\right\vert
},\\
=m^{d-1}\Omega_{d-1}\int_{0}^{\infty}\frac{x^{d-2}\sin^{2}\left(  \left(
m\Delta\right)  \sqrt{x^{2}-1}\right)  }{\left\vert x^{4}-1\right\vert }dx,
\end{array}
\end{equation}
where $x^{2}=\vec{p}^{2}/m^{2},$ while $\Omega_{d-1}$ is the volume of the
solid angle in $d-1$ dimensions, $\Omega_{2}=2\pi,\;\Omega_{3}=4\pi,$ etc..
This is a convergent integral for $d<\left(  5-\varepsilon\right)  $
dimensions, hence $n\left(  m,\Delta\right)  $ is finite for $d=1,2,3,4$
dimensions. We note that the number density $n\left(  m,\Delta\right)  $
increases monotonically at fixed $m$ as $\Delta$ increases. The energy density
per unit volume for the produced particles for all momenta is%
\begin{equation}%
\begin{array}
[c]{c}%
\rho\left(  m,\Delta\right)  =\int\frac{d^{d-1}p}{\left(  2\pi\right)  ^{d-1}%
}\sqrt{\vec{p}^{2}+m^{2}}\left\vert \beta\right\vert ^{2}\\
\;\;\;\;=\frac{m^{d}\Omega_{d-1}}{\left(  2\pi\right)  ^{d-1}}\int_{0}%
^{\infty}\frac{x^{d-2}\sqrt{x^{2}+1}\sin^{2}\left(  \left(  m\Delta\right)
\sqrt{x^{2}-1}\right)  }{\left\vert x^{4}-1\right\vert }dx
\end{array}
\end{equation}
$\rho\left(  m,\Delta\right)  $ is convergent for $d<\left(  4-\varepsilon
\right)  $ dimensions, and is logarithmically divergent at $d=4$ despite the
rapid oscillations at the ultraviolet limit.

Recall that the massive field is not a scale invariant model. In the Weyl
symmetric limit, $m\rightarrow0,$ there is no particle production at all in
any dimension. \ In the scale invariant theory masses for fields must come
from interactions, such as interactions with the Higgs field. In a
cosmological context the Higgs field is not just a constant, and therefore in
the type of investigation above, the parameter $m$ should be replaced by the
cosmological behavior of the Higgs field (see \cite{BCT-geodesics} for an
example). This very different behavior in a Weyl invariant theory should be
the more serious approach for investigating effectively massive fields to
answer the type of questions discussed in this section.

\section{Conformally exact sign-flipping backgrounds in string theory}

We consider the worldsheet formulation of the relativistic string, but we make
string theory consistent with target space Weyl symmetry as suggested in
\cite{BST-string}. This requires promoting the string tension to a dynamical
field, $\left(  2\pi\alpha^{\prime}\right)  ^{-1}\rightarrow T\left(  X^{\mu
}\left(  \tau,\sigma\right)  \right)  $. The background field $T\left(
X\right)  ,$ along with any other additional background fields, must be
restricted to satisfy exact worldsheet conformal symmetry at the quantum
level. In the worldsheet formalism, typically the tension appears together
with the metric $g_{\mu\nu}\left(  X\left(  \tau,\sigma\right)  \right)  $ or
antisymmetric tensor $b_{\mu\nu}\left(  X\left(  \tau,\sigma\right)  \right)
$ in the Weyl invariant combination, $Tg_{\mu\nu}$ or $Tb_{\mu\nu}.$ The
requirement of exact \textit{worldsheet} conformal symmetry constrains these
target-space Weyl invariant combinations. Perturbative worldsheet conformal
symmetry (vanishing beta functions) is captured by the properties of the low
energy effective string action. From the study of the Weyl invariant and
geodesically complete formalism of the low energy string action
\cite{BST-string} we have learned that the tension (closely connected to the
gravitational constant) switches sign generically near the singularities in
the classical solutions of this theory. If we fix the target space Weyl
symmetry by choosing the string gauge as in Eq.(\ref{EsGuages}), then in those
generic solutions, the tension becomes $T\left(  X^{\mu}\left(  \tau
,\sigma\right)  \right)  =\pm\left(  2\pi\alpha^{\prime}\right)  ^{-1}$ on the
two sides of the singularity as it appears \textit{in the string gauge}. Those
two sides are identified as the gravity/antigravity sectors of the low energy
theory as discussed in section (\ref{frames}). From the perspective of the
worldsheet string theory these observations lead to a simple prescription to
capture all these effects in the string gauge, namely replace the Weyl
invariant structures $\left(  Tg_{\mu\nu},Tb_{\mu\nu}\right)  $ by $\left(
\pm\left(  2\pi\alpha^{\prime}\right)  ^{-1}G_{\mu\nu}^{\pm},\pm\left(
2\pi\alpha^{\prime}\right)  ^{-1}B_{\mu\nu}^{\pm}\right)  ,$ where the capital
$\left(  G_{\mu\nu}^{\pm}\left(  X\right)  ,B_{\mu\nu}^{\pm}\left(  X\right)
\right)  $ are the background fields on the gravity/antigravity patches that
are joined at the singularities \textit{as they appear in the string gauge}.
We may absorb the overall $\pm$ due to the signs of the tension into a
redefinition of the background fields, and as we did for the Einstein gauge in
Eq.(\ref{g-hat}), define
\begin{equation}
\left(  \hat{G}_{\mu\nu}\left(  X\right)  ,\hat{B}_{\mu\nu}\left(  X\right)
\right)  =\left(  \pm G_{\mu\nu}^{\pm}\left(  X\right)  ,\pm B_{\mu\nu}^{\pm
}\left(  X\right)  \right)  ,
\end{equation}
as the full set of background fields in the union of the gravity/antigravity
sectors of the worldsheet string theory. Of course, $\left(  \hat{G}_{\mu\nu
}\left(  X\right)  ,\hat{B}_{\mu\nu}\left(  X\right)  \right)  $ are required
to satisfy worldsheet conformal invariance at the quantum level as usual. What
is new is the geodesic completeness of the background fields $\left(  \hat
{G}_{\mu\nu}\left(  X\right)  ,\hat{B}_{\mu\nu}\left(  X\right)  \right)  $
which is achieved by the sign flipping tension and the union of the
corresponding gravity/antigravity sectors.

\subsection{String in flat background with tension that flips sign
\label{flat}}

A simple example of a conformally exact worldsheet CFT, that includes a
dynamical string tension that flips signs, is the flat string background
$\eta_{\mu\nu}$ modified only by a time dependent string tension $T\left(
X\right)  =\frac{1}{2\pi\alpha^{\prime}}$Sign$\left(  \left\vert X^{0}\left(
\tau,\sigma\right)  \right\vert -\frac{\Delta}{2}\right)  .$ This can also be
presented in the string gauge by absorbing the sign of the tension into a
redefined metric%
\begin{equation}
\hat{G}_{\mu\nu}\left(  X\right)  =\eta_{\mu\nu}\text{Sign}\left(  \left\vert
X^{0}\left(  \tau,\sigma\right)  \right\vert -\frac{\Delta}{2}\right)  ,\text{
\ }\hat{B}_{\mu\nu}\left(  X\right)  =0,
\end{equation}
where $\Delta$ is a constant. Note the similarity to Eq.(\ref{Hparticle}) or
sections (\ref{NoMass},\ref{Mass}). Thus the tension is positive when
$\left\vert X^{0}\left(  \tau,\sigma\right)  \right\vert >\frac{\Delta}{2}$
and negative when $-\frac{\Delta}{2}<X^{0}\left(  \tau,\sigma\right)
<\frac{\Delta}{2}.$ This is also similar to the cosmological example with an
antigravity loop given in \cite{BST-string}, but we have greatly simplified it
here by keeping only the signs but not the magnitude of the tension, thus
defining a conformally exact rather than a conformally approximate CFT on the
worldsheet. The corresponding worldsheet string model is
\begin{equation}
S=-\frac{1}{4\pi\alpha^{\prime}}\int d^{2}\sigma\sqrt{-h}h^{ab}\partial
_{a}X^{\mu}\partial_{b}X^{\nu}\eta_{\mu\nu}\text{Sign}\left(  \left\vert
X^{0}\left(  \tau,\sigma\right)  \right\vert -\frac{\Delta}{2}\right)  .
\label{newFlatString}%
\end{equation}

We should mention that it is also possible to consider a model, at least at
the classical level, by inserting in the action (\ref{newFlatString}) the
inverse of the Sign function $\left(  \text{Sign}\left(  \left\vert
X^{0}\left(  \tau,\sigma\right)  \right\vert -\frac{\Delta}{2}\right)
\right)  ^{-1}.$ In this case the tension flips sign when it is infinite
rather than zero. Both of these possibilities occur smoothly rather than
suddenly in cosmological backgrounds in string theory (see Eq.(30) in
\cite{BST-string} or its generalizations). Both behaviors are significant from
the perspective of string theory because perturbative versus non-perturbative
methods would be needed to understand fully the physics in the vicinity of the
gravity/antigravity transitions. Namely, when the tension at the transition is
large the string would be close to being pointlike, so the stringy corrections
would be small and perturbative in the vicinity of the gravity/antigravity
transitions; by contrast when the tension at the transition is small the
string would be floppy so stringy corrections could be significant. In the
latter case, high spin fields \cite{Vasiliev} may be an interesting tool to
investigate the gravity/antigravity transition in our setting.

From the form of the action in Eq.(\ref{newFlatString}) it is evident that the
string action is invariant under reparametrizations of the worldsheet at the
classical level. We will use this symmetry to choose a gauge to perform the
classical analysis below. But eventually we also need to know if this symmetry
is valid also at the quantum level. The generator of this gauge symmetry is
the stress tensor, so the stress tensor vanishes as a constraint to impose the
gauge invariance. At the classical level the stress tensor does vanish as part
of the solution of the classical equations and constraints (see below). At the
quantum level, in \textquotedblleft`covariant quantization\textquotedblright,
the stress tensor does not vanish on all states but only on the gauge
invariant physical states. For consistency of covariant quantization one must
verify that the constraints form a set of first class constraints that close
under quantum operator products. In our case the stress tensor derived from
(\ref{newFlatString}) has the form $T_{\pm\pm}=($Sign)$\times T_{\pm\pm}^{0}$
, where $T_{\pm\pm}^{0}$ is the usual worldsheet stress tensor in the flat
background $\eta_{\mu\nu}$, while the sign factor switches signs at the kinks
$\left\vert X^{0}\left(  \tau,\sigma\right)  \right\vert =\frac{\Delta}{2}$.
In the positive (gravity, Sign=+) region, we have symbolically the operator
products, $T_{\pm\pm}^{0}\times T_{\pm\pm}^{0}\sim T_{\pm\pm}^{0},$ where the
standard CFT result on the right hand side is computed exactly for the flat
string. Similarly, in the negative (antigravity, Sign=$-$) region we have,
$\left(  -T_{\pm\pm}^{0}\right)  \times\left(  -T_{\pm\pm}^{0}\right)
\sim-\left(  -T_{\pm\pm}^{0}\right)  .$ So the algebra is closed like the
standard CFT locally in the positive and negative regions away from the kinks.
There remains analyzing the operator products at the kinks $\left\vert
X^{0}\left(  \tau,\sigma\right)  \right\vert =\frac{\Delta}{2}$ (worldsheet
analogs of the kinks in Fig.1). The operator products involving the Sign
factor non-trivially introduce delta functions and derivatives of delta
functions multiplied by the sign factor or its derivatives that have support
only at the kinks. At one contraction (order $\hbar$ effects) the coefficient
of the delta function includes the flat $T_{\pm\pm}^{0}$ or its derivatives
evaluated at the kinks. Since $T_{\pm\pm}^{0}$ or its derivatives are in the
list of first class constraints (Virasoro operators), this is still a closed
algebra of first class constraints all of which vanish on physical states. At
two contractions (order $\hbar^{2}$ in quantum effects) there are again some
terms that contain $T_{\pm\pm}^{0}$ or its derivatives, which again are of no
concern since these still vanish on physical states. However, there are also
additional operators of the form of $\left(  \partial X^{0}\right)  $
multiplying products of the sign function, delta function, or its derivatives,
all evaluated at the kinks. We have analyzed these complicated distributions
and found that they vanish when integrated with $\left(  \partial
X^{0}\right)  $, so they do not seem to contribute. Similarly we can drop
several similar terms due to the properties of the distributions. The analysis
at the kinks becomes harder at higher contractions ($\hbar^{3}$ and beyond in
quantum effects), and we leave this for future analysis to be reported at a
later stage. The main point is that if there are additional constraints that
must be imposed at the kinks they will show up in this type of operator
product analysis. So far, we have not found new constraints up to two
contractions in the operator products. Thus, the algebra of the operator
products is basically the standard algebra of a conformal field theory (CFT)
locally in the positive and negative regions away from the kinks. The
modification of the CFT algebra at the kinks with terms that are proportional
to Virasoro operators does not change the validity of the gauge symmetry at
the quantum level, since those terms vanish on physical states anyway.
Although we have not yet found other operator modifications of the algebra at
the kinks, conceptually it is possible that such terms may arise at higher
contractions or in other models that include gravity/antigravity transitions.
When and if such terms appear, they must be included in an enlarged list of
constraints that should form a closed algebra under operator products, then
this will \ define the proper quantum theory.

In this paper our aim is to first understand the classical theory of a string
described by the action in (\ref{newFlatString}), so we do not need to be
concerned here about the subtleties described in the previous paragraph. In
fact the classical analysis that we give below is helpful in further
developing the right approach for the quantum theory. Thus, setting aside
temporarily the possible stringy corrections, we are at first interested in
the classical behavior of strings as they propagate in the union of the
gravity/antigravity regions, and later try to figure out the possible
additional effects due to interactions at those transitions by using more
sophisticated methods, such as string field theory, or others, as outlined in
section \ref{sft}.

\subsubsection{General string propagating classically through antigravity}

In this section we will discuss the properties of the model in
Eq.(\ref{newFlatString}). The main objective is to show that there are no
problems due to the negative tension during antigravity from the point of view
of fundamental principles, such as unitarity or possible instability due to
negative kinetic energy. The unitarity of this string model was already
established in \cite{BST-string} more generally for any time dependent tension
$T\left(  X^{0}\right)  ,$ and more general metric, so we will not repeat it
here. We will concentrate on the effect of the antigravity period on the
propagation of the string and the corresponding signals that observers in
gravity may detect. As we will demonstrate, as compared to the complete
absence of antigravity, the presence of an antigravity period for a certain
amount of time causes only a time delay in the propagation of an open or
closed free string of any configuration. This may seem surprising since, at
first thought, one may think that string bits would fly apart under an
instability caused by a negative string tension. In fact, this does not happen
because a negative tension is simply an overall sign in the action of a free
string, and this does not change the equations of motion and constraints of a
free string during antigravity.

We work in the conformal gauge at the classical level. There is a remaining
reparametrization symmetry that permits the further choice of the following
time-like gauge%
\begin{equation}
X^{0}\left(  \tau,\sigma\right)  =\left\vert H\right\vert \tau,
\end{equation}
where $H$ is the total time dependent Hamiltonian of the string while
$\left\vert H\right\vert $ is time independent. This is similar to the
massless free field in section (\ref{NoMass}). In this gauge the remaining
degrees of freedom satisfy the following equations of motion and constraints
\begin{equation}%
\begin{array}
[c]{c}%
\left(  \partial_{\tau}^{2}-\partial_{\sigma}^{2}\right)  \vec{X}\left(
\tau,\sigma\right)  =0,\\
H^{2}=\left(  \partial_{\tau}\vec{X}\pm\partial_{\sigma}\vec{X}\right)  ^{2},
\end{array}
\end{equation}
to be solved in each time region $A,B,C$ defined by
\begin{equation}
\text{ }A:\text{\ }\tau\left\vert H\right\vert <-\Delta/2,\;B:-\Delta
/2<\tau\left\vert H\right\vert <\Delta/2,\ \;C:\tau\left\vert H\right\vert
>\Delta/2.
\end{equation}
Furthermore, the solutions for $\vec{X}_{A,B,C}\left(  \tau,\sigma\right)  $
and the canonical momenta $\vec{P}_{A,B,C}\left(  \tau,\sigma\right)
=\partial_{\tau}\vec{X}_{A,B,C}\left(  \tau,\sigma\right)  \times$Sign$\left(
\left\vert H\tau\right\vert -\frac{\Delta}{2}\right)  $ should be continuous
at the boundaries $\tau\left\vert H\right\vert =\pm\Delta/2.$ The method of
solution follows the simple model in Eq.(\ref{method}) or the massive field in
Eq.(\ref{Mass}).

We will discuss the case of an open string; the closed string is treated
similarly. The general solution in each region is given in terms of the center
of mass $\left(  \vec{q},\vec{p}\right)  $ and oscillator $\left(  \vec
{\alpha}_{n},n=\pm1,\pm2,\cdots\right)  $ degrees of freedom. The general
configuration of the string in the positive tension region $A$, at a time
$\tau<-\Delta/2,$ is a general solution $\vec{X}_{A}\left(  \tau
,\sigma\right)  $ given by
\begin{equation}
\vec{X}_{A}\left(  \tau,\sigma\right)  =\vec{q}_{0}+\vec{p}\tau+\sum
_{n=-\infty,\neq0}^{\infty}\frac{i}{n}\vec{\alpha}_{n}\cos n\sigma~e^{-in\tau
}.
\end{equation}
The time independent parameters $\left(  \vec{q}_{0},\vec{p}\right)  $ and
$\left(  \vec{\alpha}_{n},n=\pm1,\pm2,\cdots\right)  $ determine the initial
configuration of the string at the time $\tau=\tau_{0}.$ From the constraint
equations we compute the time independent $\left\vert H\right\vert $ and the
remaining constraint
\begin{equation}%
\begin{array}
[c]{c}%
\left\vert H\right\vert =\sqrt{\vec{p}^{2}+\sum_{n=1}^{\infty}\vec{\alpha
}_{-n}\cdot\vec{\alpha}_{n}}\\
0=\vec{p}\cdot\vec{\alpha}_{n}+\frac{1}{2}\sum_{m=-\infty,\neq0}^{\infty}%
\vec{\alpha}_{-m}\cdot\vec{\alpha}_{n+m}%
\end{array}
\label{constr-string}%
\end{equation}
Thus the time dependent Hamiltonian that switches sign is
\begin{equation}
H\left(  \tau\right)  =\text{Sign}\left(  \left\vert H\right\vert \left\vert
\tau\right\vert -\frac{\Delta}{2}\right)  \sqrt{\vec{p}^{2}+\sum_{n=1}%
^{\infty}\vec{\alpha}_{-n}\cdot\vec{\alpha}_{n}}.
\end{equation}
Assuming the constraints (\ref{constr-string}) are satisfied at the classical
level by some set of parameters $\left(  \vec{\alpha}_{n},\vec{p}\right)  ,$
the momentum, $\overrightarrow{P}_{A}=\overrightarrow{\dot{X}}_{A},$ in region
$A$ is%

\begin{equation}
\overrightarrow{P}_{A}\left(  \tau,\sigma\right)  =\vec{p}+\sum_{n=-\infty
,\neq0}^{\infty}\vec{\alpha}_{n}\cos n\sigma~e^{-in\tau}.
\end{equation}

In region $B,$ $-\frac{\Delta}{2}<\tau\left\vert H\right\vert <\frac{\Delta
}{2},$ the solution $\left(  \vec{X}_{B},\vec{P}_{B}\right)  $ takes the same
form as above, but with a new set of parameters $\left(  \vec{q}_{B},\vec
{p}_{B},\vec{\alpha}_{nB}\right)  .$ Note that in this region there is a
non-trivial minus sign in the relation between momentum and velocity, $\vec
{P}_{B}\left(  \tau,\sigma\right)  =-\partial_{\tau}\vec{X}_{B}\left(
\tau,\sigma\right)  .$ At the transition time, $\tau_{\ast}\equiv-\frac
{\Delta}{2\left\vert H\right\vert },$ we must match the position and momentum,
therefore $\vec{X}_{A}\left(  \tau_{\ast},\sigma\right)  =\vec{X}_{B}\left(
\tau_{\ast},\sigma\right)  $ and $\partial_{\tau}\vec{X}_{A}\left(  \tau
_{\ast},\sigma\right)  =-\partial_{\tau}\vec{X}_{B}\left(  \tau_{\ast}%
,\sigma\right)  ,$ noting the negative sign in the case of velocities. Because
the matching is for every value of $\sigma$ we find that all the parameters
$\left(  \vec{q}_{B},\vec{p}_{B},\vec{\alpha}_{nB}\right)  $ are uniquely
determined in terms of the initial parameters $\left(  \vec{q}_{0},\vec
{p},\vec{\alpha}_{n}\right)  $ in region $A$. So the solution in region $B$
is
\begin{align}
\vec{X}_{B}\left(  \tau,\sigma\right)   &  =\left(
\begin{array}
[c]{c}%
\vec{q}_{0}+\vec{p}\left(  -\tau-\frac{\Delta}{\left\vert H\right\vert
}\right)  \\
+\sum_{n=-\infty,\neq0}^{\infty}\frac{i}{n}\vec{\alpha}_{n}\cos n\sigma
~e^{-in\left(  -\tau-\frac{\Delta}{\left\vert H\right\vert }\right)  }%
\end{array}
\right)  \\
\vec{P}_{B}\left(  \tau,\sigma\right)   &  =-\overrightarrow{\dot{X}}%
_{B}\left(  \tau,\sigma\right)  =\vec{p}+\sum_{n=-\infty,\neq0}^{\infty}%
\vec{\alpha}_{n}\cos n\sigma~e^{-in\left(  -\tau-\frac{\Delta}{\left\vert
H\right\vert }\right)  }%
\end{align}
There are no new constraints beyond those that are already assumed to have
been satisfied in region $A$ by the parameters $\left(  \vec{\alpha}_{n}%
,\vec{p}\right)  $. Note the structure $(-\tau-\frac{\Delta}{\left\vert
H\right\vert })$ that indicates a backward propagation similar to Fig.1 as
$\tau$ increases beyond $\tau_{\ast}$.

At the next transition time, $\tau_{\ast\ast}\equiv+\frac{\Delta}{2\left\vert
H\right\vert },$ we must connect the solution $\left(  \vec{X}_{B},\vec{P}%
_{B}\right)  $ above to the solution $\left(  \vec{X}_{C},\vec{P}_{C}\right)
$ in region $C,$ $\tau>\tau_{\ast\ast},$ which is given in terms of a new set
of parameters $\left(  \vec{q}_{C},\vec{p}_{C},\vec{\alpha}_{nC}\right)  .$
Using the matching conditions $\vec{X}_{C}\left(  \tau_{\ast},\sigma\right)
=\vec{X}_{B}\left(  \tau_{\ast},\sigma\right)  $ and $\overrightarrow{\dot{X}%
}_{C}\left(  \tau_{\ast},\sigma\right)  =-\overrightarrow{\dot{X}}_{B}\left(
\tau_{\ast},\sigma\right)  $ that include the extra minus sign for velocities
(as discussed above), we find that $\left(  \vec{q}_{C},\vec{p}_{C}%
,\vec{\alpha}_{nC}\right)  $ are all determined again uniquely in terms of the
initial parameters $\left(  \vec{q}_{0},\vec{p},\vec{\alpha}_{n}\right)  $
introduced in region $A.$
\begin{equation}%
\begin{array}
[c]{c}%
\vec{X}_{C}\left(  \tau,\sigma\right)  =\left(
\begin{array}
[c]{c}%
\vec{q}_{0}+\vec{p}\left(  \tau-2\frac{\Delta}{\left\vert H\right\vert
}\right)  \\
+\sum_{n=-\infty,\neq0}^{\infty}\frac{i}{n}\vec{\alpha}_{n}\cos n\sigma
~e^{-in\left(  \tau-2\frac{\Delta}{\left\vert H\right\vert }\right)  }%
\end{array}
\right)  \\
\vec{P}_{C}\left(  \tau,\sigma\right)  =\vec{p}+\sum_{n=-\infty,\neq0}%
^{\infty}\vec{\alpha}_{n}\cos n\sigma~e^{-in\left(  \tau-2\frac{\Delta
}{\left\vert H\right\vert }\right)  }%
\end{array}
\end{equation}

For closed strings we find a similar result but with some additional
information for region $B.$ Namely, given some solution in region $A$ that
satisfies the string equations of motion and constraints, then the solution in
regions $B,C$ are obtained by the following substitutions of $\tau$ and
$\sigma$
\begin{equation}%
\begin{array}
[c]{l}%
\vec{X}_{B}\left(  \tau,\sigma\right)  =\vec{X}_{A}\left(  -\tau-\frac{\Delta
}{\left\vert H\right\vert },~-\sigma\right)  \\
\vec{X}_{C}\left(  \tau,\sigma\right)  =\vec{X}_{A}\left(  \tau-2\frac{\Delta
}{\left\vert H\right\vert },~\sigma\right)
\end{array}
\end{equation}
Note the extra minus sign in $\sigma\rightarrow-\sigma$ in region $B$. Namely,
for the closed string the left and right movers get scrambled during
antigravity. For the open string with Neumann boundary conditions discussed
above, the sign flip $\sigma\rightarrow-\sigma$ in region $B$ has no effect
since $\cos\left(  -n\sigma\right)  =+\cos\left(  n\sigma\right)  $, but if
the open string had Dirichlet boundary conditions then $\sin\left(
-n\sigma\right)  =-\sin\left(  n\sigma\right)  $  would induce an overall sign
flip of the oscillations during the antigravity period.

Putting it all together, we see that after the antigravity period$,$ the
emergent string experiences only a time delay $2\Delta/\left\vert H\right\vert
$ as compared to the string that propagates in the complete absence of
antigravity. This is the same conclusion that was reached for the free
particle or\ the free massless field.

\subsubsection{Rotating rod propagating through antigravity}

As a concrete example of a string configuration that satisfies all the
constraints, we present the rotating rod solution that is modified by a
tension that flips sign during antigravity as in Eq.(\ref{newFlatString}). We
begin with a straight string lying along the $\hat{x}$ axis with its center of
mass located at $\vec{q}_{0},$ as given by, $\vec{X}_{0}\left(  \sigma\right)
=\vec{q}_{0}+\hat{x}\;R_{0}\cos\sigma.$ Let this string rotate in the $\left(
\hat{x},\hat{y}\right)  $ plane and translate in the $\hat{z}$ direction as
follows
\begin{equation}
\vec{X}_{A}\left(  \tau,\sigma\right)  =\vec{q}_{0}+\hat{z}p\tau+R_{0}%
\cos\sigma~\left(  \hat{x}\cos\tau+\hat{y}\sin\tau\right)  .
\end{equation}
This satisfies the constraints in Eq.(\ref{constr-string}), since
$\partial_{\tau}\vec{X}\cdot\partial_{\sigma}\vec{X}=0,$ and gives $\left\vert
H\right\vert =\left(  p^{2}+R_{0}^{2}\right)  ^{1/2}.$ Following the steps
above we compute the matching string configuration during the antigravity
period $-\frac{\Delta}{2\left\vert H\right\vert }<\tau<\frac{\Delta
}{2\left\vert H\right\vert }$%
\begin{equation}
\vec{X}_{B}\left(  \tau,\sigma\right)  =\left(
\begin{array}
[c]{c}%
\vec{q}_{0}+\hat{z}p\left(  -\tau-\theta\right) \\
+R_{0}\cos\sigma\left(
\begin{array}
[c]{c}%
\hat{x}~\cos\left(  -\tau-\theta\right) \\
+\hat{y}~\sin\left(  -\tau-\theta\right)
\end{array}
\right)
\end{array}
\right)
\end{equation}
where $\theta=\Delta\left(  p^{2}+R_{0}^{2}\right)  ^{-1/2},$ noting that this
describes a backward propagation similar to Fig.1. Finally the matching string
configuration in the time period $\tau>\frac{\Delta}{2\left\vert H\right\vert
}$ is
\begin{equation}
\vec{X}_{C}\left(  \tau,\sigma\right)  =\left(
\begin{array}
[c]{c}%
\vec{q}_{0}+\hat{z}p\left(  \tau-2\theta\right) \\
+R_{0}\cos\sigma~\left(  \hat{x}\cos\left(  \tau-2\theta\right)  +\hat{y}%
\sin\left(  \tau-2\theta\right)  \right)
\end{array}
\right)  .
\end{equation}

As promised, as compared to the complete absence of antigravity, the presence
of an antigravity period for a certain amount of time causes only a time delay
in the propagation of a string of any configuration. The string bits of a
freely propagating string do not fly apart during antigravity when the string
tension is negative!

\subsection{2D black hole including antigravity \label{2dBH}}

Another simple example is the 2-dimensional black hole \cite{WittenBH} based
on the SL$\left(  2,R\right)  /$R gauged WZW model \cite{IB-SL2R-R}. The well
known string background metric in this case is, $ds^{2}=-2\left(  1-uv\right)
^{-1}dudv,$ with $uv<1,$ where $\left(  u,v\right)  $ are the string
coordinates $X^{\mu}\left(  \tau,\sigma\right)  $ in the Kruskal-Szekeres
basis. This space is geodesically incomplete similar to the case of the four
dimensional Schwarzschild blackhole \cite{ABJ-blackhole}.

The geodesically complete modification consists of allowing the string tension
to flip sign precisely at the singularity, namely $T\left(  X\right)  =\left(
2\pi\alpha^{\prime}\right)  ^{-1}$Sign$\left(  1-uv\right)  .$ Then the new
geodesically complete 2D-blackhole action is%
\begin{equation}
S=\frac{1}{2\pi\alpha^{\prime}}\int d^{2}\sigma\sqrt{-h}h^{ab}\frac
{\partial_{a}u\partial_{b}v}{\left\vert 1-uv\right\vert }.
\end{equation}
This differs from the old 2D black hole action by the absolute value sign, and
includes the antigravity region $uv>1$ just as the 4-dimensional case
\cite{ABJ-blackhole}. Despite the extra sign, this model is an exact CFT on
the worldsheet as can be argued in the same way following
Eq.(\ref{newFlatString}). Properties of the new 2D black hole, including the
related dilaton and all orders quantum corrections in powers of $\alpha
^{\prime},$ will be investigated in detail in a separate paper
\cite{ABJ2-2Dblackhole}.

\section{String field theory with antigravity \label{sft}}

In the neighborhood of the gravity/antigravity transition, which occurs
typically at a gravitational singularity, a proper understanding of the
physics would be incomplete without the input of quantum gravity that may
possibly contribute large quantum effects. How should we estimate the effects
of quantum gravity?

We first point out that attempting to use an effective low energy field theory
that includes higher powers of curvature, such as those computed from string
theory, is the wrong approach. Higher powers of curvature capture
approximations to quantum gravity that are valid at momenta much smaller than
the Planck scale; those cannot be used to investigate the phenomena of
interest that are at the Planck scale close to the singularity. For
investigating the gravity/antigravity transition more closely, we do not see
an alternative to using directly an appropriate theory of quantum gravity that
can incorporate the geodesically complete spacetime that includes both gravity
and antigravity regions. Hence we first need to define the proper theory of
quantum gravity that is consistent with geodesic completeness. As far as we
know this notion of quantum gravity was first considered in \cite{BST-string}.

Assuming that quantum string theory is a suitable approach to quantum gravity,
we outline here how string field theory may be modified to take into account
geodesic completeness and the presence of an antigravity sector, so that it
can be used as a proper tool to answer the relevant questions.

Open and closed string field theory (SFT) is a formalism for computing
string-string interactions, including those that involve stringy gravitons. As
in standard field theory, in principle the SFT formalism is suitable for both
perturbative and non-perturbative computations. Technically SFT is hard to
compute with, but it has the advantage of being a self consistent and
conceptually complete definition of quantum gravity and the interactions with
matter. It is therefore crucial to see how antigravity fits in SFT and
therefore how the pertinent questions involving antigravity can be addressed
in a self consistent manner.

In the context of SFT, gravitational and other backgrounds in which strings
propagate are incorporated through the BRST operator $Q$ that appears in the
quadratic part of the action \cite{WittenBH}
\begin{equation}
S_{open}=Tr\left[  \frac{1}{2}AQA+\frac{g}{3}A\star A\star A\right]  .
\label{SFTaction}%
\end{equation}
The complete SFT action must also include closed strings, $S_{closed}.$ The
supersymmetric versions of these may also be considered. Here $A\left(
X\right)  $ is the string field, the product $\star$ describes string joining
or splitting, and the BRST operator $Q$ is given by%
\begin{equation}
Q=\int d\sigma\sum_{\pm}\left\{  c_{\pm}~T_{\pm\pm}\left(  X\right)  +b_{\pm
}c_{\pm}\partial c_{\pm}\right\}  .
\end{equation}
where $\left(  b_{\pm},c_{\pm}\right)  $ are the Fadeev-Popov ghosts, which is
a device of \textquotedblleft covariant quantization\textquotedblright, while
$T_{\pm\pm}\left(  X\left(  \sigma\right)  \right)  $ is the stress tensor for
left/right moving strings, associated to any conformal field theory (CFT) on
the worldsheet that is conformally exact at the quantum level.

The gravitational and other backgrounds, including a dynamical tension that
flips signs (i.e. incorporating antigravity) of the type we discussed in the
previous sections, are included in the stress tensor $T_{\pm\pm}\left(
X\right)  .$ If these backgrounds are not geodesically complete we expect that
the SFT theory is incomplete since even at the classical level on the
worldsheet there would be string solutions that would be incomplete just like
particle geodesics that would be incomplete. Thus for a geodesically complete
SFT we need to make sure that $T_{\pm\pm}\left(  X\right)  $ belongs to a
geodesically complete worldsheet string model as described in the previous
section. Examples of such string models were provided in sections
(\ref{flat},\ref{2dBH}). Similarly one can construct many more geodesically
complete backgrounds by allowing the string tension to change sign at
singularities (and perhaps more generally) as long as the CFT conditions, that
amount to $Q^{2}=0,$ are satisfied.

If the interactions in the SFT action (\ref{SFTaction}) are neglected we do
not expect dramatic effects due to the presence of antigravity since we have
seen in the previous section the effect is only a time delay as compared to
the complete absence of antigravity (as in sections \ref{delayedParticle}%
,\ref{NoMass},\ref{flat}). By including the interactions either perturbatively
or non-perturbatively we can explore the effects of antigravity in the context
of the quantum theory. In the previous sections, we have obtained a glimpse of
the phenomena that could happen, including particle (or string) production (as
in section \ref{Mass}), excitations of various string states (as in section
\ref{interact}), and more dramatic phenomena that remain to be explored.

From the discussion in the first part of section (\ref{flat}) one may gather
that we are still in the process of addressing some technicalities in the
construction of the BRST operator $Q$ for the simple model in that section. So
we are not yet in a position to perform explicit computations, but we hope we
have provided an outline of how one may formulate an appropriate theory to
address and answer the relevant questions.

There may be alternative formalisms that could provide answers more easily
than SFT, and of course those should be explored, but the advantage of SFT for
being a conceptually complete and self consistent definition of the system,
including the presence of antigravity as outlined above, is likely to remain
as an important feature of this approach because of the overall perspective
that it provides.

\section{Comments}

We have argued that a fundamental theory that could address the physical
phenomena close to gravitational singularities, either in the form of field
theory or string theory, is unlikely to be complete without incorporating
geodesic completeness. The Weyl symmetric approach to the standard model
coupled to gravity in Eq.(\ref{action}), and the similar treatment of string
theory \cite{BST-string}, generally solves this problem and naturally requires
that antigravity regions of spacetime should appear on the other side of
gravitational singularities as integral parts of the spacetime described by a
fundamental theory. There are other views that the notion of spacetime may not
even exist at the extremes close to singularities. While acknowledging that
there may be other scenarios that are little understood at this time, we
believe that our concrete proposal merits further investigation.

While emphasizing that there are nicer Weyl gauges, we have shown how
gravitational theories and string theories can be formulated in their
traditional Einstein or string frames to include effects of a Weyl symmetry
that renders them geodesically complete. A prediction of the Weyl symmetry is
to naturally include an antigravity region of field space and spacetime that
is geodesically connected to the traditional gravity spacetime at
gravitational singularities. Precisely at the singularities that appear in the
Einstein or string frames the gravitational constant or string tension flip
sign suddenly (but smoothly in nicer Weyl gauges). As shown in section
(\ref{frames}), this sign can be absorbed into a redefinition of the metric in
the Einstein or string frame, $\hat{g}_{\mu\nu}=\pm g_{\mu\nu}^{\pm},$ where
$\hat{g}_{\mu\nu}$ describes the spacetime in the union of the gravity and
antigravity regions. This definition of the complete spacetime may then be
used to perform computations in the geodesically complete theory.

The appearance of negative kinetic energy terms for some degrees of freedom
during antigravity was a source of concern. The arguments presented here show
that this was a false alarm. We argued that unitarity is not an issue either
in gravity or antigravity and that negative energy does not imply an
instability of the theory as seen by observers in the gravity region (namely,
observers like us, analyzing the universe). We made this point by studying
many simple examples and we showed that observers in the gravity sector can
deduce the existence and at least some properties of antigravity.

We have thus eliminated the initial concerns regarding unitarity or
instability of the complete theory when there is an antigravity sector with
negative kinetic energy. We have also demonstrated that there are very
interesting physical phenomena associated with antigravity that remain to be
explored concerning fundamental physics at the extremities of spacetime. These
will have applications in cosmology as in \cite{BST-antigravity}%
\cite{BST-conf}\cite{BST-Higgs}\cite{BST-string} and black hole physics as in
\cite{ABJ-blackhole}\cite{ABJ2-2Dblackhole}.

\begin{acknowledgments}
We thank Edward Witten for encouraging us to investigate antigravity for a
single degree of freedom, Ignacio J. Araya for discussions on all topics
presented in this paper, and Neil Turok for useful remarks and comments.
\end{acknowledgments}

\end{document}